\documentclass[11pt]{article}
\usepackage[sc]{mathpazo}
\usepackage{fullpage}
\usepackage[authoryear,sectionbib,sort]{natbib}
\usepackage{graphicx}
\usepackage{adjustbox}
\usepackage[boxruled,linesnumbered]{algorithm2e}
\usepackage{multirow}
\usepackage{amsmath}
\usepackage[utf8]{inputenc}
\usepackage{lineno}
\usepackage{titlesec}
\usepackage{graphicx}
\usepackage{subcaption}
\usepackage{booktabs,multirow}
\usepackage{here}
\usepackage{url}

\usepackage{listings}
\usepackage{xcolor}

\lstdefinestyle{mystyle}{
	backgroundcolor=\color{backcolour},   
	commentstyle=\color{codegreen},
	keywordstyle=\color{magenta},
	numberstyle=\tiny\color{codegray},
	stringstyle=\color{codepurple},
	basicstyle=\ttfamily\footnotesize,
	breakatwhitespace=false,         
	breaklines=true,                 
	captionpos=b,                    
	keepspaces=true,                 
	numbers=left,                    
	numbersep=5pt,                  
	showspaces=false,                
	showstringspaces=false,
	showtabs=false,                  
	tabsize=2
}

\definecolor{codegreen}{rgb}{0,0.6,0}
\definecolor{codegray}{rgb}{0.5,0.5,0.5}
\definecolor{codepurple}{rgb}{0.58,0,0.82}
\definecolor{backcolour}{rgb}{0.95,0.95,0.92}

\titleformat{\section}[block]{\Large\bfseries\filcenter}{\thesection}{1em}{}
\titleformat{\subsection}[block]{\Large\itshape\filcenter}{\thesubsection}{1em}{}
\titleformat{\subsubsection}[block]{\large\itshape}{\thesubsubsection}{1em}{}
\titleformat{\paragraph}[runin]{\itshape}{\theparagraph}{1em}{}[. ]

\title{Pattern-Based Prediction of Population Outbreaks}

\author{Gabriel R. Palma$^{1,\ast}$ \and 
Wesley A.C. Godoy$^{2}$ \and
Eduardo Engel$^{2}$ \and
Douglas Lau$^{3}$ \and
Edgar Galvan$^{1, 4, 6}$\and
Oliver Mason$^{5}$ \and
Charles Markham$^{1, 4}$ \and
Rafael A. Moral$^{1, 5}$}

\date{}

\begin{document}

\maketitle

\noindent{} 1. Hamilton Institute, Maynooth University, Maynooth, Ireland;

\noindent{} 2. Department of Entomology and Acarology, University of S\~{a}o Paulo, Piracicaba, Brazil;

\noindent{} 3. Brazilian Agricultural Research Corporation (Embrapa Trigo), Passo Fundo, Rio Grande do Sul, Brazil;

\noindent{} 4. Department of Computer Science, Maynooth University, Maynooth, Ireland;

\noindent{} 5. Department of Mathematics and Statistics, Maynooth University, Maynooth, Ireland;

\noindent{} 6. Naturally Inspired Computation Research Group, Maynooth University, Ireland;

\noindent{} $\ast$ Corresponding author; e-mail: gabriel.palma.2022@mumail.ie


\section*{Abstract}
\begin{enumerate}
\item Insect outbreaks are biotic disturbances in forests and agroecosystems that cause economic and ecological damage. This phenomenon depends on a variety of biological and physical factors. The complexity and practical importance of the issue have made the problem of predicting outbreaks a focus of recent research.

\item Here, we propose the Pattern-Based Prediction (PBP) method for predicting population outbreaks. It is based on the Alert Zone Procedure, combined with elements from machine learning. It uses information on previous time series values that precede an outbreak event as predictors of future outbreaks, which can be useful when monitoring pest species.

\item We illustrate the methodology using simulated datasets and real time series data obtained by monitoring aphids in wheat crops in Southern Brazil. We obtained an average test accuracy of $84.6\%$ in the simulation studies implemented with stochastic models, and $95.0\%$ for predicting outbreaks using the real dataset. This shows the feasibility of the PBP method in predicting outbreaks in population dynamics.

\item We benchmarked our results against established state-of-the-art machine learning methods, namely Support Vector Machines, Deep Neural Networks, Long Short Term Memory and Random Forests. The PBP method yielded a competitive performance, associated with higher true-positive rates in most comparisons, while being able to provide interpretability rather than being a black-box method. This is an improvement over current state-of-the-art machine learning tools, especially when being used by non-specialists, such as ecologists aiming to use a quantitative approach for pest monitoring.

\item We provide open-source code to implement the PBP method in Python, through the \texttt{pypbp} package, which may be directly downloaded from the Python Package Index server or accessed through \url{https://pypbp-documentation.readthedocs.io}.
\end{enumerate}

\textit{Keywords}: Alert zone procedure, deep learning, machine learning, population dynamics, time series.

\section{Introduction} 

Automated systems for syndromic surveillance have been reported in many studies in different contexts, such as public health aiming to predict disease outbreaks and also agricultural pests \citep{Madden2003, Buckeridge2007, buntgen2020return, bright2020mapping, burkom2021electronic}. These studies have demonstrated many possibilities for predicting outbreaks based on population dynamics, sampling methods, outbreak frequency, and threshold analysis. Their results have demonstrated potential to help public health actions, based on the interpretation of results provided by these algorithms, using different data sources containing simulated and observed outbreaks \citep{Buckeridge2007, chan2021approaching}. 

Historically, the algorithms used to predict outbreaks involved classical time series methods, such as ARIMA-type models, seasonal models, and partial differential equations, among others~\citep{Buckeridge2007}. These tools positively impacted public health actions by enhancing the possibility of predicting disease outbreaks, but the applications of these methods are not restricted to this area. In quantitative ecology, these applications were expanded so that many authors started representing ecological phenomena with mathematical, statistical and machine learning methods \citep{oto, Odum2005, ross98}. One recent example is the application of supervised machine learning methods for predicting infestations of pine trees by a mountain pine beetle \citep{ramazi2021}.

Examples of these applications are the representation of biological systems and the interactions between the species, such as predator-prey, host-parasitoid and competition models \citep{Odum2005, Badkundri2019}. Among the taxonomic groups used to study these applications, insects expand the possibility of developing these methods with various models inspired by different problems. Examples are the LPA (larva-pupa-adult), host-parasitoid and other models related to herbivory \citep{oto}. These models are commonly used to study biological phenomena such as outbreaks. Additionally, researchers can implement further statistical and mathematical modelling studies with insect time series data.

Insects depend on resource availability, as demonstrated by many studies involving time series \citep{Nair2001, Nair2007, Santos2017, Lantschner2019}. The co-evolution between pests and plants reinforces this influence. It shows the complexity of this system, including biochemical strategies to avoid herbivory and the genetic plasticity of pests, enhancing their capability to obtain the necessary resources from plants \citep{Wallner1987, Nair2001, Nair2007}. Pest dynamics are also based on physical and biological conditions, such as temperature, humidity, precipitation or irrigation, which strongly influence pest density \citep{Wallner1987, Nair2001, Odum2005}. Biological factors such as mating system, life cycle, number of offspring per generation, mortality, and resource availability are also capable of influencing insect populations \citep{Wallner1987, Godfray1994, Nair2001, Hall2017}. In monoculture scenarios, crop phenology is followed by the presence of pests reinforcing that resource availability has a strong effect on the population dynamics in agroecosystems \citep{Nair2001,Nair2007,Santos2017}.

Insect outbreaks have frequently been documented in pest populations \citep{Santos2017, Lynch2009, Lynch2018}. They are important biotic disturbances in forests and agroecosystems \citep{Wallner1987, Nair2001, Lantschner2019}, since they may cause economic and/or ecological damage. Studies have shown possible explanations for outbreaks, such as abundant resources in monocultures, absence of predators or parasitoids, genetic factors, and pheromones produced by pests \citep{Hall2017, Tao2012}. Biotic disturbances can be intensified by climate change and human activities, which makes it essential to study how they influence pest outbreaks \citep{Volney2000, Sharma2018, Phophi2019}. One example of human activities affecting outbreaks is the occurrence of bark beetles in temperate forests. These insects have devastated a large area of pine trees in the continental United States \citep{Negron2008}.

In Brazil, there are several examples of pest outbreaks in forests and crops, as for example \textit{Eucalyptus} with \textit{Thyrinteina arnobia} and \textit{Stenalcidia sp} (Geometridae) \citep{Zanuncio2006}; black wattle with \textit{Oncideres impluviata} (Cerambycidae) showing annual outbreaks in the state of Rio Grande do Sul \citep{Ono2014}; soybean with \textit{Chrysodeixis includes} and \textit{Anticarsia gemmatalis} (Lepidoptera: Noctuidae) also exhibiting high frequencies \citep{Bueno2010,Santos2017}. Studies focused on the species mentioned above show that outbreaks occur suddenly because of different natural effects that can result in increased population densities \citep{Nair2001, Nair2007, Santos2017, Lantschner2019}. However, the exact reason why an insect species population suddenly increases in number is still an open question \citep{Ekholm2019}. Outbreak forecasting turns out to be an arduous task requiring a large amount of man-hours, extensive field work and different types of specialised equipment. Entomologists traditionally have been using different interventions to reduce economic damage, mainly in agriculture. The most common method traditionally used for this task is to define an economic threshold level \citep{Stern1959, Onstad1987}.

The economic threshold is based on the fact that pest density must be considered in the economic context in which it occurs \citep{Mitchell2014}. One of the first concepts of economic threshold was published by \cite{Stern1959} which is defined as ``\textit{the density at which control measures should be determined to prevent an increasing pest population from reaching the economic-injury level}''. This concept used the term \textit{economic-injury level} which was defined in the same paper as ``\textit{the lowest population density that will cause economic damage}''. Therefore, both concepts help in decision making to determine the moment of action, which reduces the pest population density.

Nowadays, the possibilities of actions against insect damage in crops can be found in the integrated pest management domain, which briefly consists of employing biological, physical, chemical and genetic approaches to reduce the population densities of a pest \citep{Stern1959, Goodell2009}. The economic threshold concept has been coupled with more complete analyses involving control functions given by the crop and pest population information \citep{Mitchell2004, Dun2009, Tinsley2013}. When outbreaks are frequently recorded in insect populations, the probability that their population size is bigger than the economic-injury level increases, causing secondary outbreaks \citep{Goodell2009}. Given that this biological disturbance can occur suddenly and pest monitoring can be delayed, this density can be bigger than the economic threshold in the subsequent monitoring sample. 

Different approaches have been proposed to address problems of this nature. One example is the Alert Zone Procedure (AZP) \citep{Hilker2007}, which consists of scanning observations preceding population outbreak events to obtain profiles associated with these outbreaks. This method can extract meaningful information about outbreaks because previous densities before this event allow for the comprehension of ecological patterns. This method can be used as a basis to improve pest outbreak forecasting. However, it must be improved to deal with real-world problems. The effectiveness of these approaches remains an open question that is the focus of this research.

This paper proposes the Pattern-Based Prediction (PBP) method, which is an extension of the AZP based on statistical machine learning. We begin by describing the method and then carry out simulation studies to assess the performance of our method under different conditions. Finally, we illustrate our proposed method using a dataset obtained from a pest management system aimed at monitoring aphids, which are important pests for many different cultures -- such as wheat, barley, and mustard \citep{kranti2021} -- and discuss the feasibility of applying it to the context of pest management.

\section{Methods}

\subsection{Generating patterns}

\begin{figure}[htb]
	\centering
	\includegraphics[width=16cm]{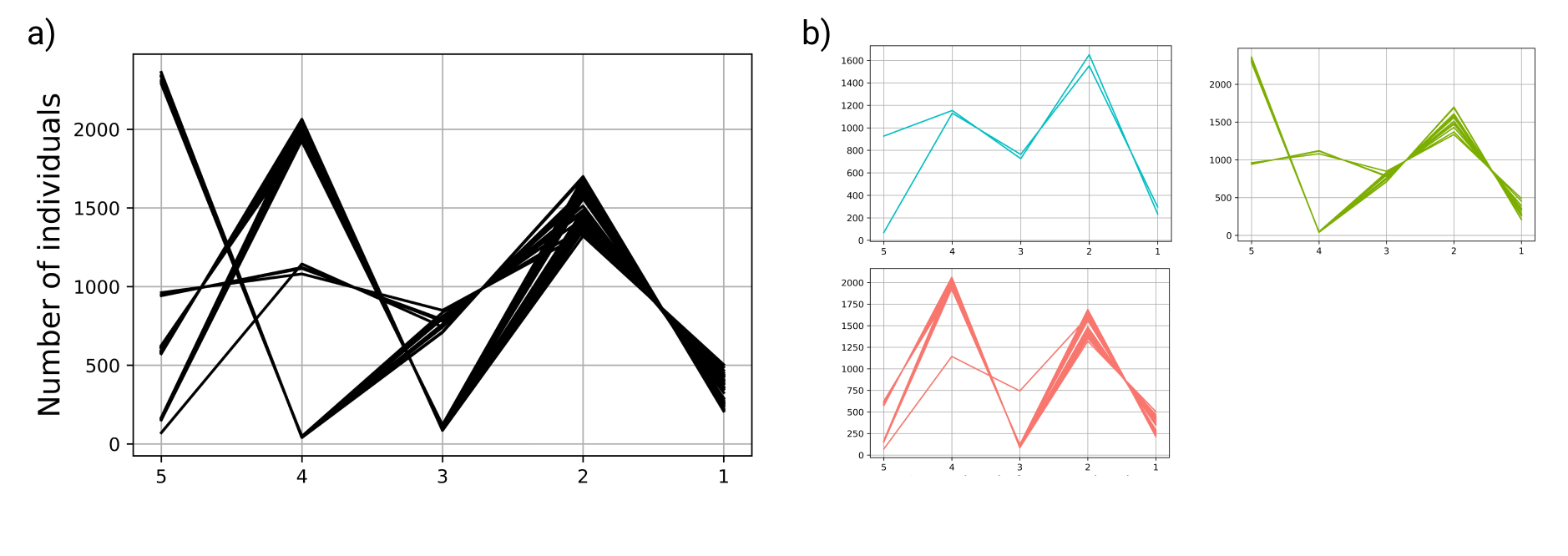}
	\caption{a) The representation of patterns $\mathbf{p}_i$ within the matrix $\mathbf{P}$ that precede an outbreak event, using $m=5$. b) The respective cluster matrices $\mathbf{P}^{\prime}_c$ obtained using $d^*_{\mbox{\scriptsize cluster}}=0.4$. These patterns were obtained from time series data simulated from a Ricker map, with $r=3$ and $K=1000$, $x_1=200$ and $1000$ observations. The population size threshold for the outbreak event was set as $x^*=2224$ representing the 90\% percentile of the data.}
	\label{fig:patterns}	
\end{figure}

Let $x_t$ represent the population size of a particular species at time point $t$, $t=1,\ldots,T$. Initially, we set a population size threshold $x^*$ such that when $x_t\geq x^*$ we have a population outbreak at time $t$. We then implement the AZP, as proposed by \cite{Hilker2007}. This method consists of scanning observations to identify each outbreak event $i$, $i = 1, \ldots, I$, that occurred at time point $t_i$, based on the value of $x^*$, and collecting the $m$ observations that precede them, forming a vector $\textbf{p}_i^T = \left\lbrace p_{i1}, p_{i2}, ..., p_{im}\right\rbrace = \left\lbrace x_{t_i-1}, x_{t_i - 2}, ..., x_{t_i-m}\right\rbrace$ per event. If $t_i-m <1$, event $i$ is ignored. After that, we group all population dynamics patterns that precede these events as the matrix

\begin{equation}
	\textbf{P}  = \begin{bmatrix}
	x_{t_1-1} & x_{t_1 - 2} & \cdots & x_{t_1-m}\\
	x_{t_2-1} & x_{t_2 - 2} & \cdots & x_{t_2-m}\\
	\vdots & \vdots & \ddots & \vdots\\
	x_{t_I-1} & x_{t_I - 2} & \cdots & x_{t_I-m}\\
	
	\end{bmatrix}= \begin{bmatrix}
	\textbf{p}^T_1\\
	\textbf{p}^T_2\\
	\vdots\\
	\textbf{p}^T_I
	
	\end{bmatrix} ,
\end{equation}\\
where $I$ represents the total number of identified patterns. See Figure~\ref{fig:patterns}(a) for a plot of all rows of a hypothetical $\mathbf{P}$ matrix. Note that time series pre-processing may be carried out prior to obtaining the pattern matrix $P$. For instance, in Section~\ref{simulationresults} we compare the performance of our method using the raw time series and a pre-processed series using the Empirical Mode Decomposition method \citep{kim2012}.

Then, if patterns $i$ and $j$ ($i\neq j$) are sufficiently similar, we group them in the same cluster. We do this based on the association metric

\begin{equation}
	d(\textbf{p}_i, \textbf{p}_{j}) =\frac{1}{c(\textbf{p}_i,\textbf{p}_j) + 1},
\end{equation}

\noindent where $c(\textbf{p}_i,\textbf{p}_j)=\sum_{k=1}^m\frac{|p_{ik} - p_{jk}|}{|p_{ik}|+|p_{jk}|}>0$ is the Canberra distance \citep{androutsos1998distance, ehsani2020robust} between two vectors, where $|\cdot|$ is the Euclidean norm. This distance is appropriate for non-negative count data \citep{androutsos1998distance}. Note that when $c(\textbf{p}_i,\textbf{p}_j)\rightarrow\infty$, then $d(\textbf{p}_i,\textbf{p}_j)\rightarrow0$, and as $c(\textbf{p}_i,\textbf{p}_j)\rightarrow0$, then $d(\textbf{p}_i,\textbf{p}_j)\rightarrow1$.

To define the similarity of patterns we set the value $d^{*}_{\mbox{\scriptsize cluster}}$, representing the minimum association metric for considering $\textbf{p}_{i}$ similar to $\textbf{p}_{j}$. This yields the cluster matrices $\mathbf{P}^{\prime}_{c}$, $c~=~1,\ldots,C$, that include patterns which are similar to one another. To obtain these, we start with pattern $\textbf{p}_{1}$, which represents the first row of the matrix $\textbf{P}$. We remove $\textbf{p}_{1}$ from $\textbf{P}$ and add it as the first row of $\textbf{P}^{\prime}_{1}$. After that we compute the association metric between $\textbf{p}_1$ and all subsequent rows of $\textbf{P}$. If $d(\textbf{p}_1, \textbf{p}_{j}) \geq d^*_{\mbox{\scriptsize cluster}}$, we add pattern $\textbf{p}_{j}$ as the last row of the cluster matrix $\textbf{P}^{\prime}_{1}$ and delete it from $\textbf{P}$. We repeat this process to obtain the cluster matrices $\textbf{P}^{\prime}_{c}, c=1,\ldots,C\leq I,$ until there are no more rows left in $\textbf{P}$ (Algorithm~1 in the supplementary material). Note that the order of $\textbf{P}$ is important for the clustering procedure. See Figure~\ref{fig:patterns}(b) for an example where the generated patterns were split into $C=4$ cluster matrices.

After obtaining the $C$ cluster matrices
\begin{equation}
\textbf{P}^{\prime}_{c}  = \begin{bmatrix}
x_{t_1-1} & x_{t_1 - 2} & \cdots & x_{t_1-m}\\
x_{t_2-1} & x_{t_2 - 2} & \cdots & x_{t_2-m}\\
\vdots & \vdots & \ddots & \vdots\\
x_{t_{l^\prime_c}-1} & x_{t_{l^\prime_c} - 2} & \cdots & x_{t_{l^\prime_c}-m}\\

\end{bmatrix},
\end{equation}
where $l^\prime_c$ is the number of rows of $\mathbf{P}^\prime_c$, we compute the vectors of means $\overline{\mathbf{p}}^{\prime}_c$, containing the mean of each column for cluster matrix $\mathbf{P}^{\prime}_c$, to form the rows of the matrix

\begin{equation}
\textbf{P}^{\prime}_{{\mbox{\scriptsize means}}}  = \begin{bmatrix}
\displaystyle\frac{1}{l^\prime_1}\sum\limits_{c=1}^{l^\prime_1}x_{t_c-1} & \displaystyle\frac{1}{l^\prime_1}\sum\limits_{c=1}^{l^\prime_1}x_{t_c - 2} & \cdots & \displaystyle\frac{1}{l^\prime_1}\sum\limits_{c=1}^{l^\prime_1}x_{t_c-m}\\
\displaystyle\frac{1}{l^\prime_2}\sum\limits_{c=1}^{l^\prime_2}x_{t_c-1} & \displaystyle\frac{1}{l^\prime_2}\sum\limits_{c=1}^{l^\prime_2}x_{t_c - 2} & \cdots & \displaystyle\frac{1}{l^\prime_2}\sum\limits_{c=1}^{l^\prime_2}x_{t_c-m}\\
\vdots & \vdots & \ddots & \vdots\\
\displaystyle\frac{1}{l^\prime_C}\sum\limits_{c=1}^{l^\prime_C}x_{t_{c}-1} & \displaystyle\frac{1}{l^\prime_C}\sum\limits_{c=1}^{l^\prime_C}x_{t_{c} - 2} & \cdots & \displaystyle\frac{1}{l^\prime_C}\sum\limits_{c=1}^{l^\prime_C}x_{t_{c}-m}\\

\end{bmatrix}= \begin{bmatrix}
\overline{\textbf{p}}^{\prime T}_1\\
\overline{\textbf{p}}^{\prime T}_2\\
\vdots\\
\overline{\textbf{p}}^{\prime T}_{C}

\end{bmatrix} .
\end{equation}

\begin{figure}[htb]
	\centering
	
	\includegraphics[width=10cm]{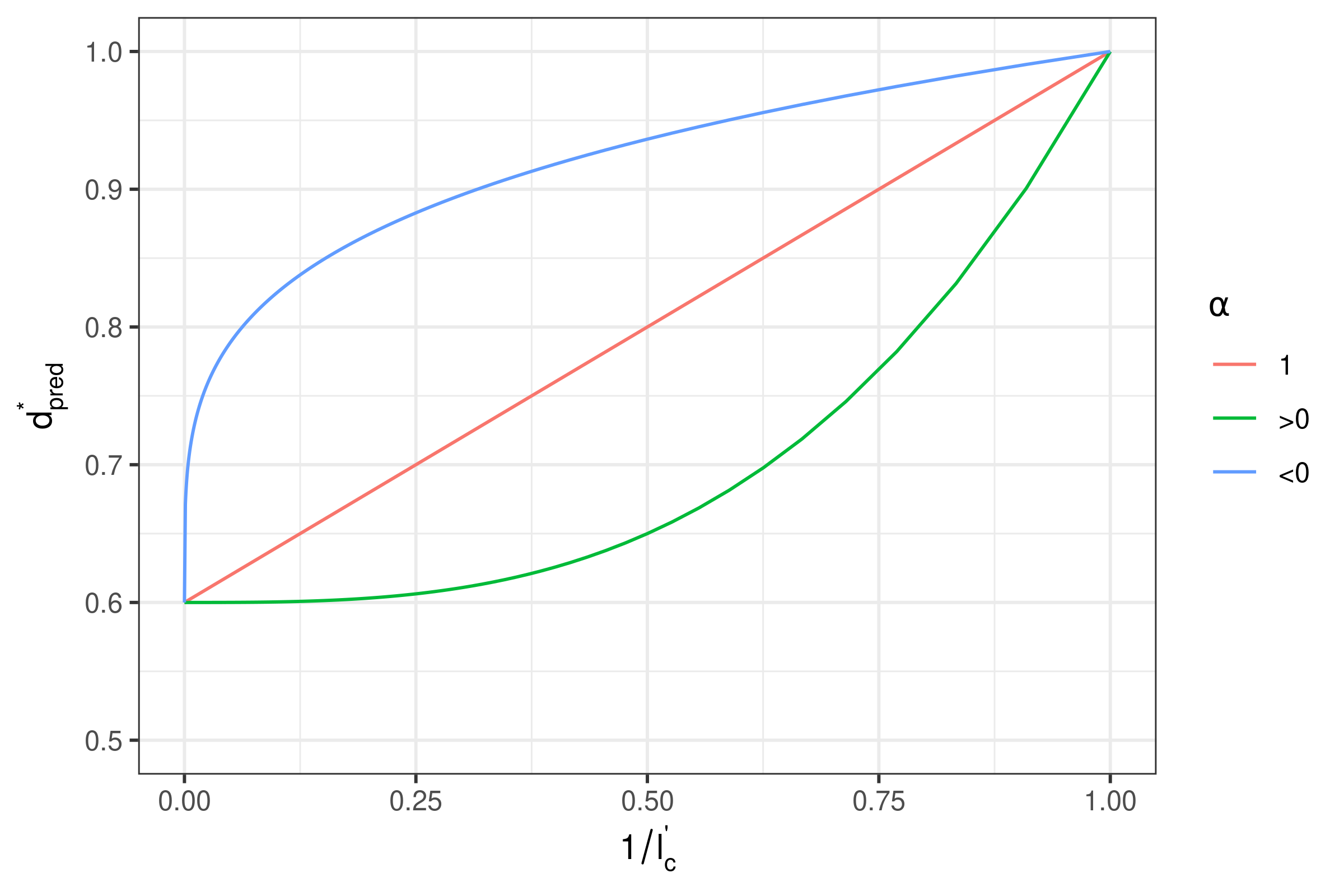}
	\caption{The threshold for prediction $d^{*}_{\mbox{\scriptsize pred}}$, calculated as a function of $l^\prime_c$ for $\alpha=1$ (red curve), $\alpha=3$ (green curve) and $\alpha=0.25$ (blue curve), whilst fixing $d^*_{\mbox{\scriptsize base}}=0.6$. The $x$-axis is represented as $1/l^\prime_c$ to ease visualisation.}
	\label{fig:alpha}
	
\end{figure}

The matrix $\textbf{P}^{\prime}_{{\mbox{\scriptsize means}}}$ contains the information of all cluster matrices $\textbf{P}^{\prime}_{c}$, and is used for the prediction of a future event. Given a new collection of observations $\textbf{x}_{\mbox{\scriptsize new}}$, with length $m$, we compute the association metric between $\textbf{x}_{\mbox{\scriptsize new}}$ and each row of $\textbf{P}^{\prime}_{{\mbox{\scriptsize means}}}$.  If any computed association is greater or equal to $d^{*}_{\mbox{\scriptsize pred}}$, shown in Eq.~\ref{d_pred_eq}, the threshold for prediction, we predict that an event will occur. Finally $d^{*}_{\mbox{\scriptsize pred}}$ is defined for each row of $\textbf{P}^{\prime}_{{\mbox{\scriptsize means}}}$ as a function of $l^\prime_c$, the number of patterns that generated each vector of means:
\begin{equation}
d^*_{\mbox{\scriptsize pred}}=f(l^\prime_c) = d^{*}_{\mbox{\scriptsize base}} +  \frac{(1 - d^{*}_{\mbox{\scriptsize base}})}{(l^\prime_c)^\alpha},
\label{d_pred_eq}
\end{equation}
where $d^{*}_{\mbox{\scriptsize base}}$ is the baseline value of the association metric (the smallest it is allowed to be) and $\alpha$ is a constant that changes the shape of the function $f$ (see Figure~\ref{fig:alpha}). When $l^\prime_c\rightarrow \infty$, we have that $d^*_{\mbox{\scriptsize pred}}\rightarrow d^{*}_{\mbox{\scriptsize base}}$, and as $l^\prime_c\rightarrow1$, also $d^*_{\mbox{\scriptsize pred}}\rightarrow1$. This means that for predicting that a new event will occur, we would need a larger association between $\mathbf{x}_{\mbox{\scriptsize new}}$ and a particular $\overline{\mathbf{p}}^\prime_c$ that was obtained from a small number of patterns.

\begin{figure}[htb]
	\centering
	
	\includegraphics[width=10cm]{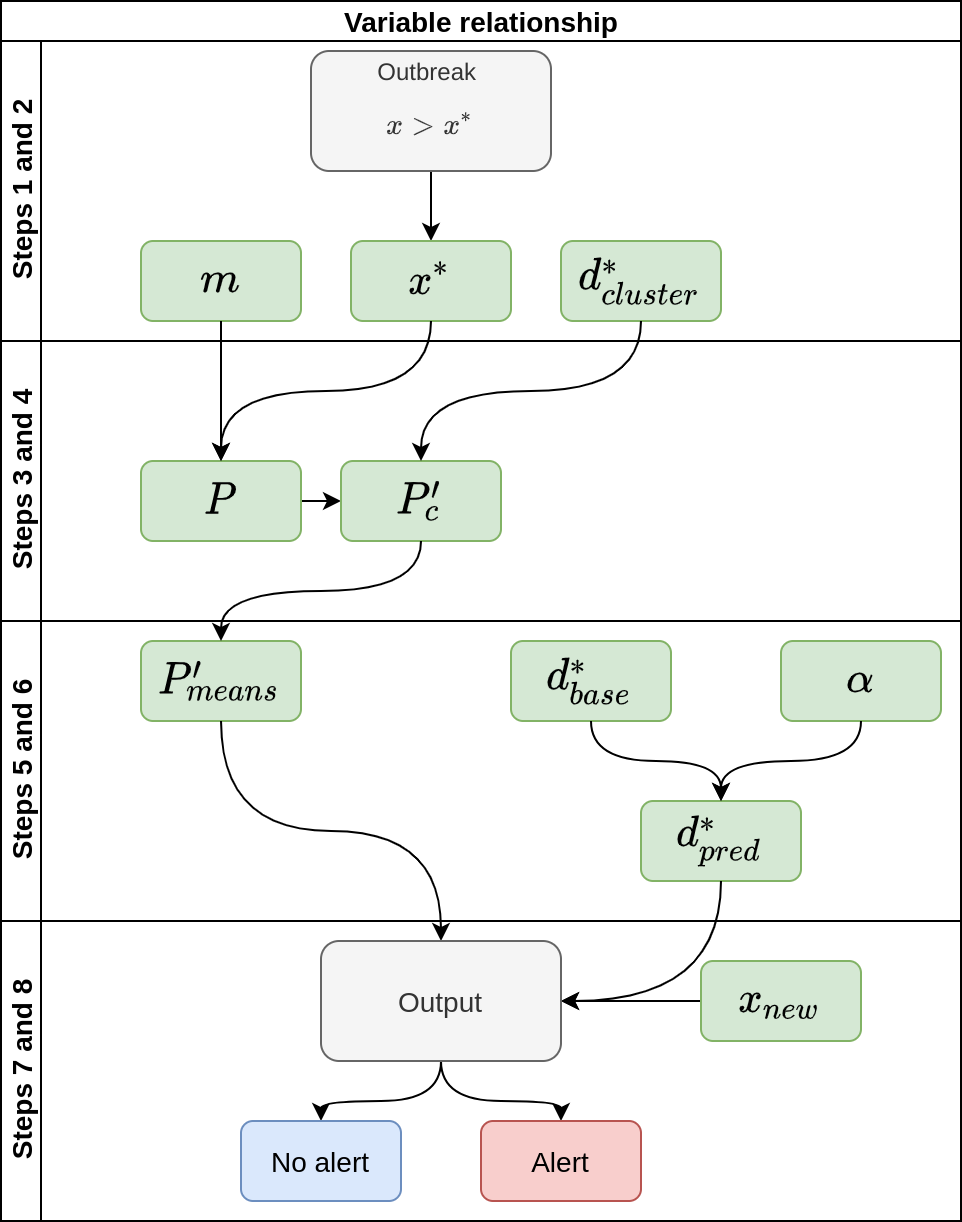}
	\caption{A schematic representation of the pattern-based method used to predict an outbreak based on time series data.}
	\label{fig:scheme}
	
\end{figure}

To summarise, the PBP method consists of the following steps:
\begin{enumerate}
	\item Choose the value of the population size threshold $x^{*}$;
	\item Set the values of $m$ and $d^{*}_{\mbox{\scriptsize cluster}}$;
	\item Generate the pattern matrix $\mathbf{P}$;
	\item Obtain the cluster matrices $\mathbf{P}^\prime_c$ (algorithm presented in the supplementary material);
	\item Compute the matrix $\textbf{P}^{\prime}_{{\mbox{\scriptsize means}}}$ from the column means of each cluster matrix $\mathbf{P}^\prime_c$;
	\item Set the values of $d^{*}_{\mbox{\scriptsize base}}$ and $\alpha$ and obtain $d^*_{\mbox{\scriptsize pred}}$ for each row of $\textbf{P}^{\prime}_{{\mbox{\scriptsize means}}}$;
	\item Given a new collection of observations $\textbf{x}_{\mbox{\scriptsize new}}$ compute the proposed association metric between $\textbf{x}_{\mbox{\scriptsize new}}$ and each row of $\textbf{P}^{\prime}_{{\mbox{\scriptsize means}}}$;
	\item If the computed association coefficient is greater than or equal to the value of $d^*_{\mbox{\scriptsize pred}}$ associated with that row of $\textbf{P}^{\prime}_{{\mbox{\scriptsize means}}}$, predict that a new event will occur at the next time step; predict that it will not occur, otherwise.
\end{enumerate}
A schematic diagram of this process is represented in Figure~\ref{fig:scheme}.

\subsection{Choosing $m$, $d^{*}_{\mbox{\scriptsize cluster}}$, $d^{*}_{\mbox{\scriptsize base}}$ and $\alpha$ via cross-validation}

We propose the use of k-fold cross-validation to choose the values of $m$, $d^{*}_{\mbox{\scriptsize cluster}}$ and $\alpha$, such that the accuracy of the method is optimized. Here, the k-fold cross validation briefly consists of creating $k$ groups of patterns $\mathbf{p}_i$ of the pattern matrix $\mathbf{P}$ and by removing the first group of patterns from $\mathbf{P}$, obtaining $\textbf{P}^{\prime}_{{\mbox{\scriptsize means}}}$ without using the information of this group, and carrying out the method to predict the occurrence of events based on the first group. After obtaining the method predictions based on this group, we compute \begin{itemize}
	\item true positives (TP): the number of times the method accurately predicted an event;
	\item true negatives (TN): the number of times the method accurately predicted there was no event;
	\item false positives (FP): the number of times the method incorrectly predicted an event;
	\item and false negatives (FN): the number of times the method incorrectly predicted there was no event.
\end{itemize} We use these values to obtain the accuracy, $\mbox{ACC} = \frac{\mbox{TP}+\mbox{TN}}{\mbox{TP}+\mbox{TN}+\mbox{FP}+\mbox{FN}}$, the true positive rate, $\mbox{TPR} = \frac{\mbox{TP}}{\mbox{TP}+\mbox{FN}}$, and the false positive rate, $\mbox{FPR}=\frac{\mbox{FP}}{\mbox{TN}+\mbox{FP}}$. We repeat this process for each group ending up with $k$ values of these metrics. To measure the overall performance, we obtain the average of these metrics. Here, we carry out the analysis using $k=5$.

To optimise the predictive power of the method, firstly we fix the values of $m$, $d^*_{\mbox{\scriptsize cluster}}$ and $\alpha$, and obtain different TPR and FPR values by varying $d^*_{\mbox{\scriptsize base}}$. The TPR and FPR can be plotted against each other to form a ROC curve \citep{Hastie2004}. This curve is bounded between 0 and 1. For a method with good predictive power, we expect the area under the curve (AUROC) to be close to 1. Let $g(m, d^*_{\mbox{\scriptsize cluster}}, \alpha)$ be an objective function that returns $-\mbox{AUROC}$ based on the described method. Then, we use the Generalized Simulated Annealing method \citep{tsallis1988, tsallis1996, xiang1997, xiang2000, xiang2013, mullen2014} to obtain the values of $m$, $d^*_{\mbox{\scriptsize cluster}}$ and $\alpha$ that minimise $g$. This method speeds up the computation process compared to a grid search of the variables. Other methods may be used, such as Differential Evolution (see the supplementary materials).

To construct the ROC curve, we vary $d^{*}_{\mbox{\scriptsize base}}$ from $0$ to $1$ using increments of $0.1$ and calculate the AUROC using the trapezoid method \citep{LIU1994}. Finally, we apply one of two forms to choose $d^{*}_{\mbox{\scriptsize base}}$: the first is based on selecting a minimum threshold for the true positive rate (e.g. 0.8 or 0.9) and the second on selecting a maximum threshold for the false positive rate (e.g. 0.1 or 0.2). In summary, the method consists of the following steps:
\begin{enumerate}
	\item Fix the values of $m$, $d^*_{\mbox{\scriptsize cluster}}$ and $\alpha$;
	\item For different values of $d^*_{\mbox{\scriptsize base}}$, carry out $k$-fold cross validation and obtain the TPR and FPR for each fold, and compute the AUROC;
	\item Choose $m$, $d^*_{\mbox{\scriptsize cluster}}$ and $\alpha$ such that the AUROC is the largest;
	\item Choose $d^*_{\mbox{\scriptsize base}}$ based on the minimum TPR or maximum FPR that would be allowed in the study, based on the ROC curve with the largest area. (Note that the minimum TPR or maximum FPR allowed depends highly on the ecological system and objectives of the monitoring programme.)
\end{enumerate}

\subsection{Sensitivity analysis}

We carried out a sensitivity analysis using simulated deterministic population dynamics, obtained from the Ricker map \citep{oto}:
\begin{equation}
	x_{t+1} = x_{t}\exp{\left[r\left( 1 - \frac{x_{t}}{K}\right) \right]},
\end{equation}  
where $x_{t}$ denotes the population size of an organism at time $t$ and the parameters $r > 0$ and $K>0$ describe the intrinsic growth rate and carrying capacity of the environment, respectively. We simulated $100$ generations using $r = 3$, $K =1000$ and an initial value of $x_1=200$.

In order to study the influence of different values of $m$ and $d^{*}_{\mbox{\scriptsize cluster}}$ on $C$ (the total number of cluster matrices $\mathbf{P}^\prime$) and the overall accuracy of the method, we used the simulated observations from the Ricker map, setting $x^*=2224$ as the threshold for an outbreak event. This value corresponds to the 90\% percentile of the simulated values from the deterministic Ricker map setting the parameter values as described above. We then employed the methodology described in the previous sections to obtain the matrices $\textbf{P}^{\prime}_{{\mbox{\scriptsize means}}}$ for $m$ varying from 2 to 15 in increments of 1, and $d^{*}_{cluster}$ varying from 0 to 1 with increments of $0.1$.

\subsection{Method validation under stochastic conditions}

To study the accuracy of the proposed method in predicting outbreak and extinction risk events under stochastic conditions, we simulated from three different approaches. The first included an additive Gaussian error $\varepsilon_{t}\sim \mbox{Normal}(0,\sigma^{2})$ in the Ricker map, yielding the recurrence equation
\begin{equation}
x_{t+1} = x_{t}\exp{\left[r\left( 1 - \frac{x_{t}}{K}\right) \right]} + \varepsilon_{t+1}.
\label{Normal}
\end{equation}
Whenever the addition of the random noise term yielded $x_{t+1}<0$, a new random noise value would be drawn from the normal distribution until $x_{t+1}>0$, to ensure positive population sizes.

The second approach utilized a state-space formulation using a Poisson distribution with the mean term $\mu$ given by the Ricker recurrence equation, i.e.
\begin{eqnarray}
X_1 &\sim &\mbox{Poisson}(\mu_1=x_1) \\
X_{t+1}|X_t &\sim & \mbox{Poisson}\left(\mu_{t+1}=X_t\exp\left[r\left(1-\frac{X_t}{K}\right)\right]\right),
\end{eqnarray}
from which all $x_t$ values were drawn recursively. Finally, the third approach utilizes a state-space formulation based on a negative binomial distribution to accommodate overdispersion in the simulation study, i.e. $X_{t+1}|X_t\sim\mbox{Negative Binomial}(\mu_{t+1},\phi)$. We estimated $r$ and $K$ based on real time series data of aphid counts in Southern Brazil for each model formulation, as well as the dispersion parameters $\sigma^2$ for the Gaussian model and $\phi$ for the negative binomial model.

Using the parameter estimates in Table~\ref{tab:estimates}, we simulated $20$ samples of size $400$ for each model. We also simulated $20$ samples of size $400$ using the negative binomial model with $\phi=3$, to introduce a scenario with stronger overdispersion. We computed the accuracy, TPR and FPR by training the methods with the initial $80\%$ observations and testing with $20\%$ of the time series. Moreover, based on the ROC curve with the largest AUROC, we chose $d^*_{\mbox{\scriptsize base}}$ using four methods:
\begin{enumerate}
    \item `TPR\_08': choose the $d^*_{\mbox{\scriptsize base}}$ value associated with the smallest TPR value that is equal to or greater than 0.8;
    \item `TPR\_09': choose the $d^*_{\mbox{\scriptsize base}}$ value associated with the smallest TPR value that is equal to or greater than 0.9;
    \item `FPR\_01': choose the $d^*_{\mbox{\scriptsize base}}$ value associated with the largest FPR value that is equal to or less than 0.1;
    \item `FPR\_02': choose the $d^*_{\mbox{\scriptsize base}}$ value associated with the largest FPR value that is equal to or less than 0.2.
\end{enumerate}

\begin{table}[htb]
    \centering
    \begin{tabular}{cp{1.5cm}p{1.5cm}p{1.5cm}}
    \hline
        \multirow{2}{*}{Parameter} & \multicolumn{3}{c}{Model} \\ \cmidrule{2-4}
         & \multicolumn{1}{c}{Gaussian} & \multicolumn{1}{c}{Poisson} &
        \multicolumn{1}{c}{Negbin}\\ \hline
        $r$ & \multicolumn{1}{r}{0.15} & \multicolumn{1}{r}{0.28}& \multicolumn{1}{r}{0.57}\\
        $K$ & \multicolumn{1}{r}{224}& \multicolumn{1}{r}{310} & \multicolumn{1}{r}{370}\\
        $\sigma^2$ & \multicolumn{1}{r}{$21,866$} & \multicolumn{1}{r}{$-$}&\multicolumn{1}{r}{$-$} \\
        $\phi$ & \multicolumn{1}{r}{$-$}&\multicolumn{1}{r}{$-$} & \multicolumn{1}{r}{1.2} \\
        \hline
        AIC & \multicolumn{1}{r}{$5,226$}& \multicolumn{1}{r}{$34,913$}&\multicolumn{1}{r}{$4,296$} \\
        \hline
    \end{tabular}
    \caption{Parameter estimates obtained when fitting the Ricker state-space model to the aphid data assuming different distributions for the observation process, namely Gaussian, Poisson and negative binomial, as well as the Akaike Information Criterion (AIC) for each model fit. Negbin = negative binomial.}
    \label{tab:estimates}
\end{table}

\subsection{Analysis of case-study: Aphid data}

To illustrate the predictive performance of our method, we use data obtained from an aphid monitoring programme implemented in Southern Brazil (State of Rio Grande do Sul, RS). These insects are considered as important pest species of many crops. For instance, among the aphid species monitored by this programme, the species \textit{Rhopalosiphum padi} and \textit{Rhopalosiphum rufiabdominalis} are widely considered important pest species associated with winter cereals, and are found in the Eurasian region with a cosmopolitan distribution \citep{Macfadyen2012}. Sampling was carried out weekly in an area of $5500 m^2$ in a wheat culture region (Coxilha, RS, 710 m altitude, $28^\circ 11^\prime 42.8^{\prime\prime}$ S and $52^\circ 19^\prime 30.6^{\prime\prime}$ W), from 2011 to 2019, totalling 424 observations. The temperature and relative humidity data were monitored at the Passo Fundo weather station ($28^\circ 15^\prime$ S, $52^\circ 24^\prime$ W, 684 m), located 10 km from the experimental area. The field was cultivated under a no-till system.

The species of aphids were monitored using Moericke traps (yellow tray, 45 cm long x 30 cm wide x 4.5 cm high), filled with a solution (2 L) consisting of water, 40\% formalin (0, 3\%) and detergent (0.2\%). Each tray had three lateral holes (5 mm in diameter) close to the edge, protected by a thin screen to prevent leaks and loss of solid content during rain. Four traps were distributed at the borders of the crop rotation tests. The traps were levelled at approximately 20 cm from the floor with bricks. The crop rotation area was cultivated with cereals (oat, wheat, and triticale), radish and fallow during the Winter and in the Summer with soybeans, corn, and \textit{Brachiaria} sp. Every seven days, the solid content of the trays was separated from the solution through the sieve and collected. The biological material was preserved in a glass bottle with 70\% alcohol. Aphids and parasitoids were separated, identified, and counted under a stereomicroscope in the laboratory.

Monitoring is one of the bases for Integrated Pest Management (IPM). For aphids, their importance stands out mainly due to the ability of these insects to transmit viruses (Barley Yellow Dwarf Virus) to economically important crops, such as wheat, oats, barley and rye. The criterion to find a threshold ($x^*$) was the total number of aphids observed in the four traps. Usually, $10\%$ of plants infested by aphids results in an economic threshold \citep{bell2015long}. This percentage corresponds to 50 insects per trap, totalling 200 aphids, which is the threshold value used here to define an outbreak.

We selected 40\%, 50\%, 60\% and 70\% of the initial observations of the time series for training and the complement as test sets to obtain the accuracy, true-positive rates and false-positive rates. We compared the performance of the PBP method with Random Forests (RF), Support Vector Machines (SVM), Deep Neural Networks (DNNs) and Long Short Term Memory (LSTM) algorithms. To obtain the algorithm performance using these competing methodologies, we created the matrix \textbf{P} with $m = 1$, 4 and 7. Another matrix that did not contain outbreaks was generated with the same $m$ observations before a threshold lower than the population size $x^* = 200$ that defines an outbreak of aphids in the study area.

For RF, we used 2 splitting predictors per tree (for $m=4$ and 7), and a total of 1,000 trees. For SVM, we used the linear kernel. For DNN, after experimenting with different architectures, we used 17 hidden layers with 11, 12, 13, 14, 15, 14, 13, 12, 11, 10, 9, 8, 7, 6, 5, 4, and 3 neurons each, all using the rectified linear unit (ReLU) activation function; for the output layer we used the sigmoid activation function. We used the LSTM method by adding a long short term memory block on the DNN's architecture previously described. The implementation of the classification methods was carried out in Python using the libraries TensorFlow and Scikit-learn \citep{python}. Finally, we obtained the accuracy, true positive rate and false positive rate using the aforementioned training-test splits of the time series.

\section{Results and discussion}

\subsection{Sensitivity analysis}

\begin{figure}[htb]
	\centering
	\begin{subfigure}{.45\textwidth}
		\centering
		\includegraphics[width=1.1\linewidth]{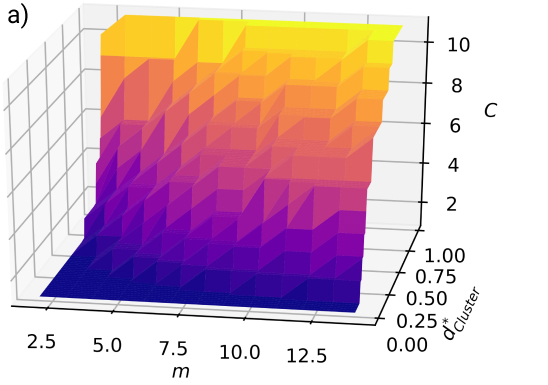}
		\label{fig:asd}
	\end{subfigure}
	\begin{subfigure}{.45\textwidth}
		\centering
		\includegraphics[width=1.1\linewidth]{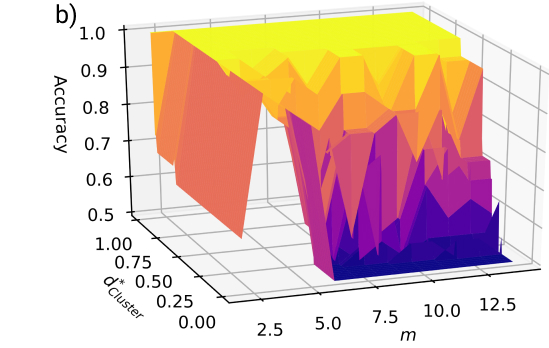}
		\label{fig:sfig2}
		
	\end{subfigure}

	\caption{The effect of $m$ and $d^{*}_{cluster}$ on (a) $C$ (i.e. number of cluster-matrices $\textbf{P}^{\prime}_{c}$), and on (b) the accuracy of the proposed method. These results were obtained from time series data simulated from a Ricker map, with $r=3$ and $K=1000$, with $x_1=200$. The population size threshold for the outbreak event was set as $x^*=2224$ representing the 90\% percentile of the data.}
	\label{Sensitivity} 
	
\end{figure}

We found that $d^{*}_{cluster}$ is proportional to the number of cluster matrices ($C$) created by the proposed method (see Figure~\ref{Sensitivity}(a)). However, as $d^{*}_{cluster}$ reaches values higher than $0.45$, the parameter $m$ did not influence the accuracy of our methods. It indicates that fixed values of $m$ could be used when we use such values of $d^{*}_{cluster}$ (Figure~\ref{Sensitivity}(b)). These findings highlight the importance of using optimisation procedures to choose the appropriate value of $d^{*}_{cluster}$ for each study.

The creation of cluster matrices and subsequent sensitivity analysis carried out here is an increment to the studies conducted by \cite{Hilker2007}. By using the AZP as a basis to create $\textbf{P}^{\prime}_{c}$ we enhance the information which can be extracted from the population dynamics of any species of interest. This process can extract the different population states using the dynamics obtained from monitoring programmes. Also, by grouping these states into different cluster matrices, we can observe the frequency of each patterns group occurring before the outbreak. The process of clustering will help us to perform the outbreak classification based on the different pattern numbers contained in $\textbf{P}^{\prime}_{c}$. This result reflects the accuracy of $100\%$ obtained in some parameter regions of the sensitivity study. As shown in the following sections, the results are dramatically improved when we fully optimise our choice for the parameter values, even when subject to stochastic effects.

\subsection{Method validation under stochastic conditions}

\begin{figure}[!ht]
	\centering
	
	\includegraphics[width=17cm]{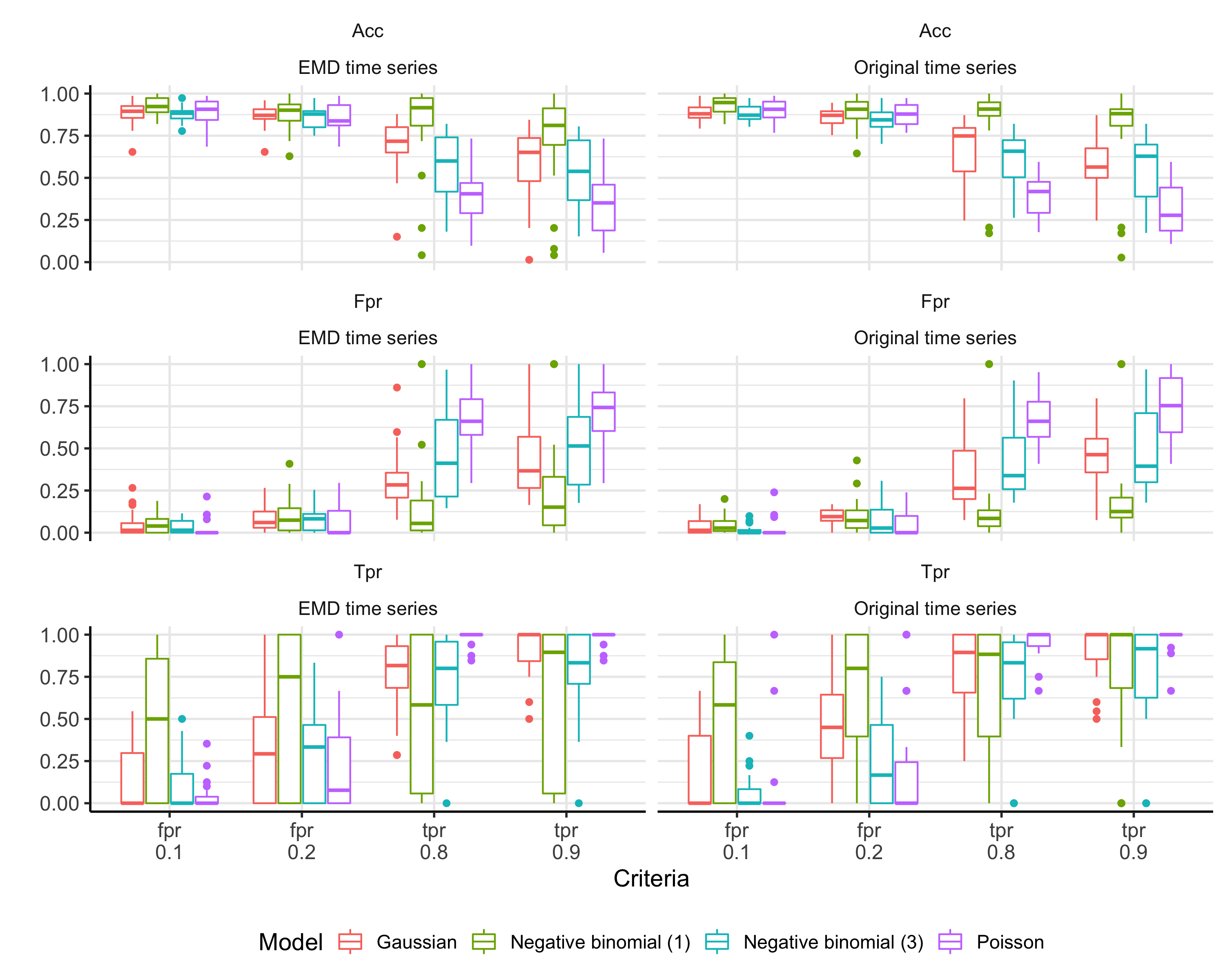}
	\caption{Accuracy simulation results, TPR (True Positive Rate) and FPR (False Positive Rate) using the raw simulated time series and pre-processed series using Empirical Mode Decomposition (EMD). In both scenarios four methods were used to choose $d^*_{\mbox{\scriptsize base}}$: based on a maximum FPR (0.1 and 0.2) or a minimum TPR (0.8 and 0.9).}
	\label{simulationresults}
	
\end{figure}

The accuracy of our method for predicting population outbreaks obtained from the stochastic simulation scenarios using the raw time series data and pre-processing the data using Empirical Mode Decomposition (EMD) considering all models were, respectively, on average $73.8\%$ with a standard deviation of $23.4\%$ and $73.2\%$ with a standard deviation of $24.1\%$ (see Figure~\ref{simulationresults}). The average FPR obtained was $25.3\%$ with a standard deviation of $29.0\%$ and $25.1\%$ with a standard deviation of $28.7\%$. Finally, the average TPR were $53.2\%$ with a standard deviation of $40.9\%$ and $55.6\%$ with a standard deviation of $41.9\%$. Therefore, we found that there are no differences in performance when pre-processing the data using EMD.

Also, considering the influence of the model, we found that the rank of models in which our method produced higher performance is, respectively, starting with the best one, the negative binomial ($\phi = 1.2$), Gaussian, negative binomial ($\phi = 3$) and Poisson stochastic models. On average, we obtain an accuracy of $84.6\%$ with a standard deviation of $20.5\%$, a false positive rate of $14.9\%$ with a standard deviation of $24.2\%$ and a true positive rate of $59.6\%$ with a standard deviation of $42.0\%$ for the negative binomial ($\phi = 1.2$) model. Considering the Gaussian model, we obtained an accuracy of $75.2\%$ with a standard deviation of $18.0\%$, a false positive rate of $22.9\%$ with a standard deviation of $22.2\%$ and a true positive rate of $55.5\%$ with a standard deviation of $37.3\%$. This finding makes our method a promising prediction tool since we got good results even when using stochastic approaches to simulate the data.

\subsection{Analysis of case-study: Aphid data}

\begin{figure}[htb]
	\centering
	
	\includegraphics[width=13cm]{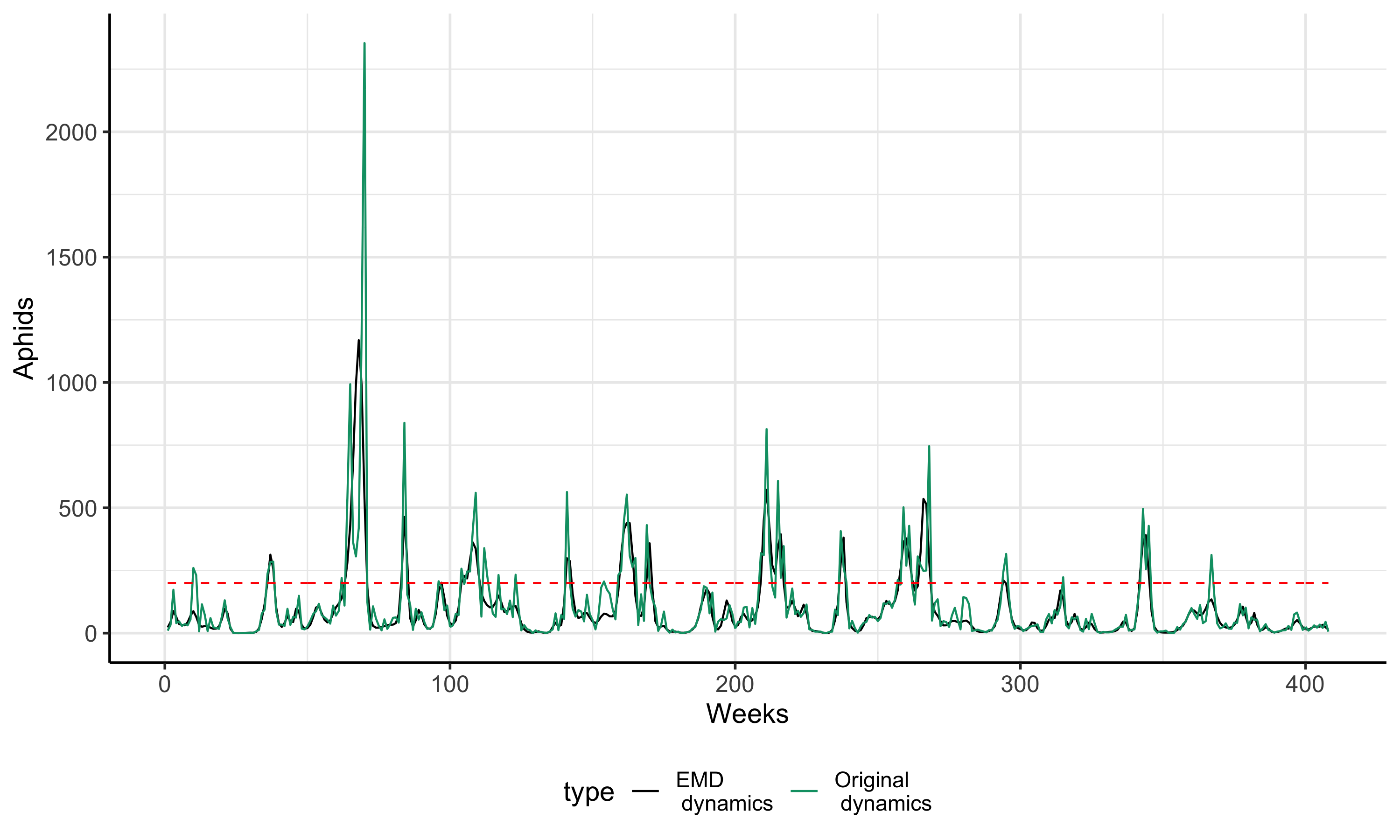}
	
	\caption{The time series represents the total aphids collected within the four traps on the monitoring system on time. The red line represents the threshold $x^{*} = 200$, the green line is the original time series, and the black line is the result of the empirical mode decomposition method.}
	\label{aphidDinamics}
	
\end{figure}

To predict the threshold representing an outbreak for the aphid population dynamics (Figure~\ref{aphidDinamics}), we select $x^*=200$ considering the number of species collected in the four traps of the monitoring system, which was related to the $10\%$ of plants infected by aphids resulting in the economic threshold. Applying PBP using different training data obtained from percentages of the initial observations of the time series of aphids, the accuracy values were higher than $70\%$ regardless of the percentage of training using the original aphid time series.

\begin{table}[htb]
	\caption{ Prediction accuracy, true-positive rate (TPR) and false positive rate (FPR) obtained from the Pattern-Based Prediction (PBP) and competing methods Random Forests (RF), Support Vector Machines (SVM), Deep Neural Networks (DNNs) and Long Short Term Memory (LSTM) methods with different values of $m$ (i.e. observations before the event). All methods were carried out using training sets with $ 40\%, 50\%, 60\%, 70\%$ and $80\%$ of the initial observation of the aphid time series. }
	\label{methodComparison}
	\centering
	\begin{adjustbox}{max width=\textwidth}
	\begin{tabular}{crrrrrrrrrrrrrrrrr}
		\hline
		
		Metrics&
		\multicolumn{1}{c}{Train}& 
		\multicolumn{1}{c}{PBP} & 
		\multicolumn{1}{c}{PBP}&
		\multicolumn{1}{c}{PBP}&
		\multicolumn{1}{c}{PBP} &
		\multicolumn{1}{c}{RF}&
		\multicolumn{1}{c}{RF}&
		\multicolumn{1}{c}{RF}&
		\multicolumn{1}{c}{SVM}&
		\multicolumn{1}{c}{SVM}&
		\multicolumn{1}{c}{SVM}&
		\multicolumn{1}{c}{DNN}&
		\multicolumn{1}{c}{DNN}&
		\multicolumn{1}{c}{DNN}&
		\multicolumn{1}{c}{LSTM}&
		\multicolumn{1}{c}{LSTM}&
		\multicolumn{1}{c}{LSTM}
		\\[-0.3cm]
		&
		\multicolumn{1}{c}{percentage}& 
		\multicolumn{1}{c}{($FPR = 0.1$)} & 
		\multicolumn{1}{c}{($FPR = 0.2$)}&
		\multicolumn{1}{c}{($TPR = 0.8$)} &
		\multicolumn{1}{c}{($TPR = 0.9$)}&
		\multicolumn{1}{c}{$(m = 1)$}&
		\multicolumn{1}{c}{$(m = 4)$}&
		\multicolumn{1}{c}{$(m = 7)$}&
		\multicolumn{1}{c}{$(m = 1)$}&
		\multicolumn{1}{c}{$(m = 4)$}&
		\multicolumn{1}{c}{$(m = 7)$}&
		\multicolumn{1}{c}{$(m = 1)$}&
		\multicolumn{1}{c}{$(m = 4)$}&
		\multicolumn{1}{c}{$(m = 7)$}&
		\multicolumn{1}{c}{$(m = 1)$}&
		\multicolumn{1}{c}{$(m = 4)$}&
		\multicolumn{1}{c}{$(m = 7)$}
		\\
		\hline
     
    & 0.4 & 0.81 & 0.81 & 0.71 & 0.49 & 0.86 & 0.91 & 0.90 & 0.89 & 0.89 & 0.89 & 0.90 & 0.88 & 0.93& 0.89 & 0.89&0.94\\[-0.3cm] 
  & 0.5 & 0.87 & 0.87 & 0.87 & 0.87 & 0.87 & 0.91 & 0.90 & 0.87 & 0.87 & 0.88 & 0.89 & 0.87 & 0.93& 0.87& 0.88&0.94\\[-0.3cm]  
  Accuracy & 0.6 & 0.90 & 0.88 & 0.83 & 0.83 & 0.89 & 0.92 & 0.90 & 0.91 & 0.89 & 0.89 & 0.90 & 0.89 & 0.93 & 0.90&0.89&0.94\\[-0.3cm]  
  &0.7 & 0.94 & 0.94 & 0.92 & 0.87 & 0.92 & 0.95 & 0.95 & 0.95 & 0.94 & 0.94 & 0.92 & 0.94 & 0.98 & 0.94&0.94&0.96\\[-0.3cm]  
  &0.8 & 0.95 & 0.90 & 0.90 & 0.90 & 0.95 & 0.95 & 0.96 & 0.96 & 0.95 & 0.94 & 0.94 & 0.95 & 0.98 & 0.95&0.94&0.99\\[-0.1cm]  
  
  &0.4 & 0.52 & 0.52 & 0.78 & 0.96 & 0.42 & 0.60 & 0.55 & 0.07 & 0.04 & 0.03 & 0.41 & 0.00 & 0.34&0.00&0.11&0.53\\[-0.3cm] 
  &0.5 & 0.18 & 0.18 & 0.18 & 0.18 & 0.50 & 0.68 & 0.60 & 0.04 & 0.04 & 0.00 & 0.42 & 0.00 & 0.36&0.00&0.12&0.64\\[-0.3cm] 
  TPR&0.6 & 0.12 & 0.69 & 0.87 & 0.87 & 0.44 & 0.60 & 0.55 & 0.12 & 0.00& 0.00 & 0.31 & 0.00 & 0.31&0.00&0.06&0.56\\[-0.3cm] 
  &0.7 & 0.28 & 0.28 & 0.43 & 0.71 & 0.29 & 0.60 & 0.60 & 0.14 & 0.00& 0.00 & 0.14 & 0.00 & 0.71&0.00&0.00&0.43\\[-0.3cm] 
  &0.8 & 0.50 & 0.75 & 0.75 & 0.75 & 0.50 & 0.50 & 0.67 & 0.25 & 0.00& 0.00 & 0.25 & 0.00 & 0.75&0.00&0.00&0.75\\[-0.1cm]
  
  &0.4 & 0.15 & 0.15 & 0.30 & 0.57 & 0.10 & 0.05 & 0.04 & 0.00 & 0.00& 0.00 & 0.04 & 0.00 & 0.00& 0.00& 0.00& 0.01\\[-0.3cm] 
  &0.5 & 0.05 & 0.05 & 0.05 & 0.05 & 0.07 & 0.04 & 0.03 & 0.00 & 0.00& 0.00 & 0.04 & 0.00 & 0.00& 0.00& 0.00& 0.02\\[-0.3cm] 
  FPR & 0.6 & 0.01 & 0.09 & 0.18 & 0.18 & 0.07 & 0.04 & 0.04 & 0.00 & 0.01& 0.01 & 0.03 & 0.00 & 0.00& 0.00& 0.01& 0.01\\[-0.3cm] 
  &0.7 & 0.02 & 0.02 & 0.05 & 0.11 & 0.04 & 0.02 & 0.02 & 0.00 & 0.00& 0.00 & 0.03 & 0.00& 0.00& 0.00& 0.00& 0.00\\[-0.3cm] 
  &0.8 & 0.03 & 0.09 & 0.09 & 0.09 & 0.01 & 0.03 & 0.01 & 0.00 & 0.00& 0.00 & 0.03 & 0.00& 0.00& 0.00& 0.00& 0.00\\[0.2cm] 
   \hline
	\end{tabular}
	\end{adjustbox}
	
\end{table}

Table~\ref{methodComparison} shows that the PBP method is competitive with state-of-the-art machine learning methods, such as the commonly used Random Forest (RF) algorithm. The criterion of a false positive rate of at most $0.2$ provides an accuracy of $91.0\%$, a false positive rate of $8\%$ and a true positive rate of $75\%$. Only DNN could obtain similar values of true positive rate for this case study. Moreover, our method using the criteria based on the true positive rate values of a minimum of $0.8$ and $0.9$ could obtain higher values of $96\%$. On the other hand, the method got an accuracy of $71\%$ and a false positive rate of $32\%$ in both cases. It indicates the flexibility of our approach because users can allow for a trade-off between false positives and false negatives. In practice, a false positive would result in using pest control techniques unnecessarily, while a false negative could result in failing to control the pest, which may cause economic damage due to an outbreak occurring.

\begin{figure}[htb]
	\centering
	
	\includegraphics[width=13cm]{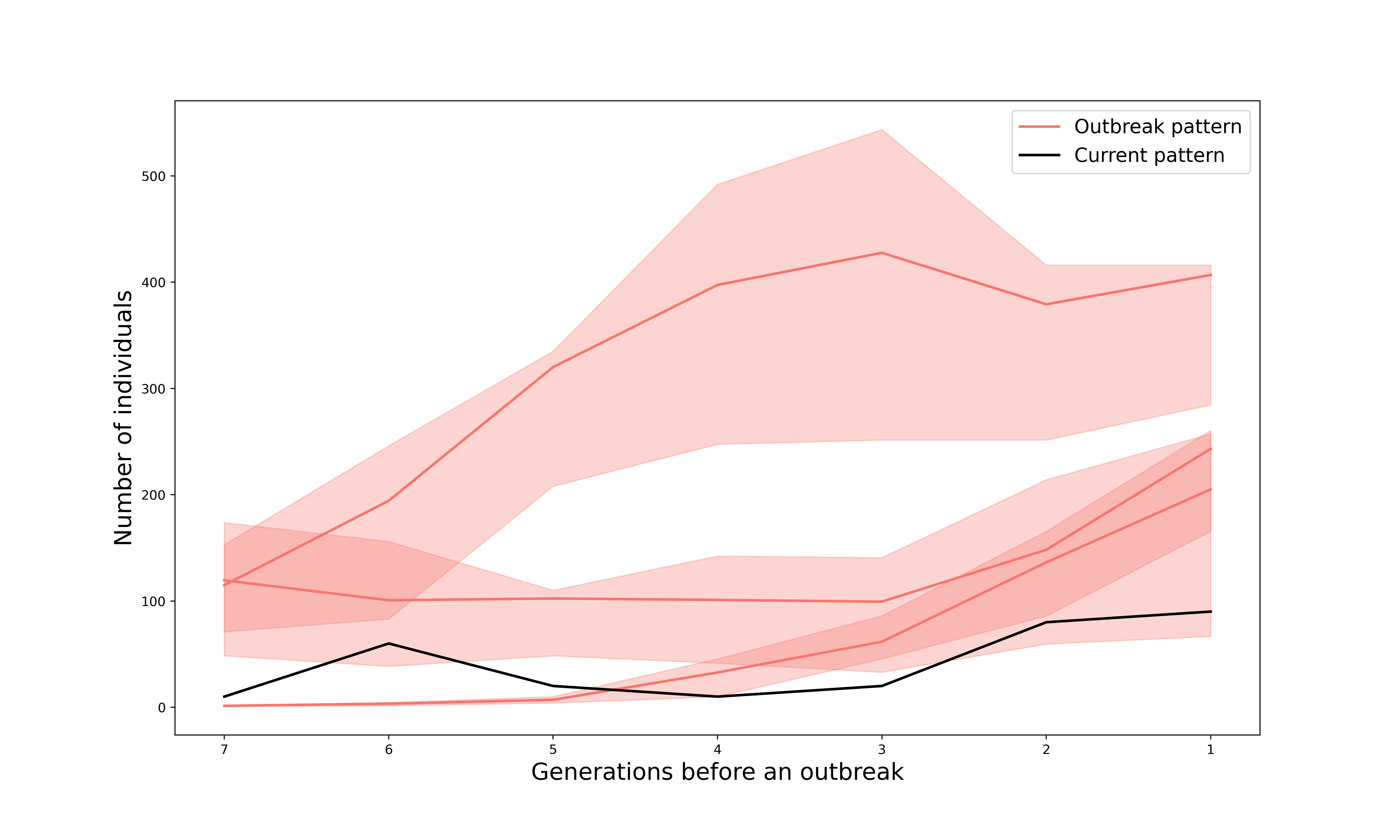}
	
	\caption{Outbreak patterns obtained from the proposed method using the aphid time series. Each of the red lines represents a row of the $\textbf{P}^{\prime}_{{\mbox{\scriptsize means}}}$ matrix. The intervals are the $25\%$ and $75\%$ percentiles of the patterns that generated each vector of means. The black line represents an observed series ($\textbf{x}_{\mbox{\scriptsize new}}$), for which the association metric $d$ is calculated between each of the three identified patterns. If $d>d^{*}_{\mbox{\scriptsize pred}}$, then the PBP method would classify $\textbf{x}_{\mbox{\scriptsize new}}$ as preceding an outbreak event. The calculated $d^{*}_{\mbox{\scriptsize pred}}$ values for the three patterns were $0.37$, $0.61$ and $0.42$, whereas the association metrics between $\textbf{x}_{\mbox{\scriptsize new}}$ and each pattern were $0.20$, $0.26$ and $0.15$, respectively. Therefore, $\textbf{x}_{\mbox{\scriptsize new}}$ would be classified as not preceding an outbreak.}
	\label{pypbpplot}
	
\end{figure}
With respect to interpretability, when using RF, it is possible to obtain variable importance. However in this case, this will tell us which previous steps were most important when predicting outbreaks, not necessarily how they relate to its occurrence. When using LSTM and DNN, which are commonly referred to as `black-box' methods \citep{liang2021}, it is even more challenging to find explainable frameworks that allow us to study the relationship between the predictors and the outbreaks. However, each hyperparameter in the PBP framework provides a clear interpretation, and we are able to create visual representations of the patterns that occurred before the outbreak (the $\textbf{P}^{\prime}_{{\mbox{\scriptsize means}}}$ matrix). For instance, Figure~\ref{pypbpplot} displays the three patterns in $\textbf{P}^{\prime}_{{\mbox{\scriptsize means}}}$ obtained from employing PBP using the optimised hyperparameter values for the aphid data using 50\% of the time series for training.

Additionally, the number of patterns encountered in each cluster matrix presents the importance of each clustered pattern for predicting animal outbreaks. The parameter $\alpha$ shows how relevant each group of patterns is for predicting an outbreak. Also, we can assess the importance of clustering the pattern matrix by looking at the estimate of $d^*_{\mbox{\scriptsize cluster}}$. Larger values typically indicate fewer recognised patterns in $\textbf{P}^{\prime}_{{\mbox{\scriptsize means}}}$. The $m$ hyperparameter shows the number of previous observations required to provide a classification based on our method, so it provides a clear interpretation for ecologists and farmers in terms of how far in the past to watch for when identifying patterns. The $d^{*}_{\mbox{\scriptsize base}}$ hyperparameter informs the minimum degree of similarity that is required to classify an outbreak, based on previously observed patterns. Therefore, not only is the PBP method competitive when compared to state-of-the-art machine learning methods, it is also interpretable, and brings descriptive advantages combined with its predictive power.



\section*{Acknowledgments}

This publication has emanated from research conducted with the financial support of Funda\c{c}\~{a}o de Amparo \`{a} Pesquisa do Estado de S\~{a}o Paulo (proc. no. 19/14805-7 and no. 20/06147-7), Ag\^{e}ncia USP de Inova\c{c}\~{a}o and Science Foundation Ireland under Grant 18/CRT/6049. The opinions, findings and conclusions or recommendations expressed in this material are those of the authors and do not necessarily reflect the views of the funding agencies.

\bibliographystyle{apalike}
\bibliography{ref}

\newpage

\section*{Supplementary Materials: Pattern-Based Prediction of Population Outbreaks}
We introduce the \texttt{pypbp} package, which is a Python implementation of the Pattern-Based Prediction (PBP) method, and examples of how to use it in Section~\ref{PBP_package}. Also, we present the clustering algorithm used by the PBP method in Section~\ref{algorithm}. Finally, we describe the methods used for the PBP optimisation procedure, including a comparison between the performance of Generalised Simulated Annealing (GSA) and differential evolution in Section~\ref{optimisation_methods}.

\section{The \texttt{pypbp} package \label{PBP_package}}
To use our package, first install Python 3 using the official website (\url{https://www.python.org}). We recommend the Jupyter Lab environment (\url{https://jupyter.org}) as a GUI for Python, however there are many other options. Using the \texttt{pip} command in your terminal (\texttt{cmd} in Windows, or \texttt{terminal} in Mac and Linux operating systems), you may install Jupyter Lab by executing
\begin{lstlisting}[language=Bash, alsoletter=.]
pip install jupyterlab
\end{lstlisting}
To install the \texttt{pypbp} package, execute
\begin{lstlisting}[language=Bash, alsoletter=.]
pip install pypbp
\end{lstlisting}
You may then open a Jupyter Lab environment using the command
\begin{lstlisting}[language=Bash, alsoletter=.]
jupyter lab --core-mode
\end{lstlisting}
Finally, you may create a new Jupyter notebook file using the Jupyter Lab environment. Using the cells of the Jupyter notebook file, you can import and use functions implemented in the \texttt{pypbp} package, as presented below. More information is available in the package description page (\url{https://pypbp-documentation.readthedocs.io/en/latest/}) and GitHub repository (\url{https://github.com/GabrielRPalma/PyPBP}). 

\subsection{A minimal reproducible example}

The following code obtains estimates for the hyperparameters in the PBP method based on the aphid data used as motivation in the paper.
\begin{lstlisting}[language=Python, alsoletter=.]
import pypbp as pbp 

results, xstar, clustered_patterns, parameters = pbp.pbp_fit(time_series = pbp.time_series, 
            train_percentage = 0.5, 
            xstar = 200, 
            maxfun = 200)
pbp.pbp_plot(time_series = pbp.time_series, clustered_patterns = clustered_patterns, 
             parameters = parameters, 
             xnew = [10, 60, 20, 10, 20, 80, 90])
\end{lstlisting}
\begin{figure}[!ht]
	\centering
	
	\includegraphics[width=13cm]{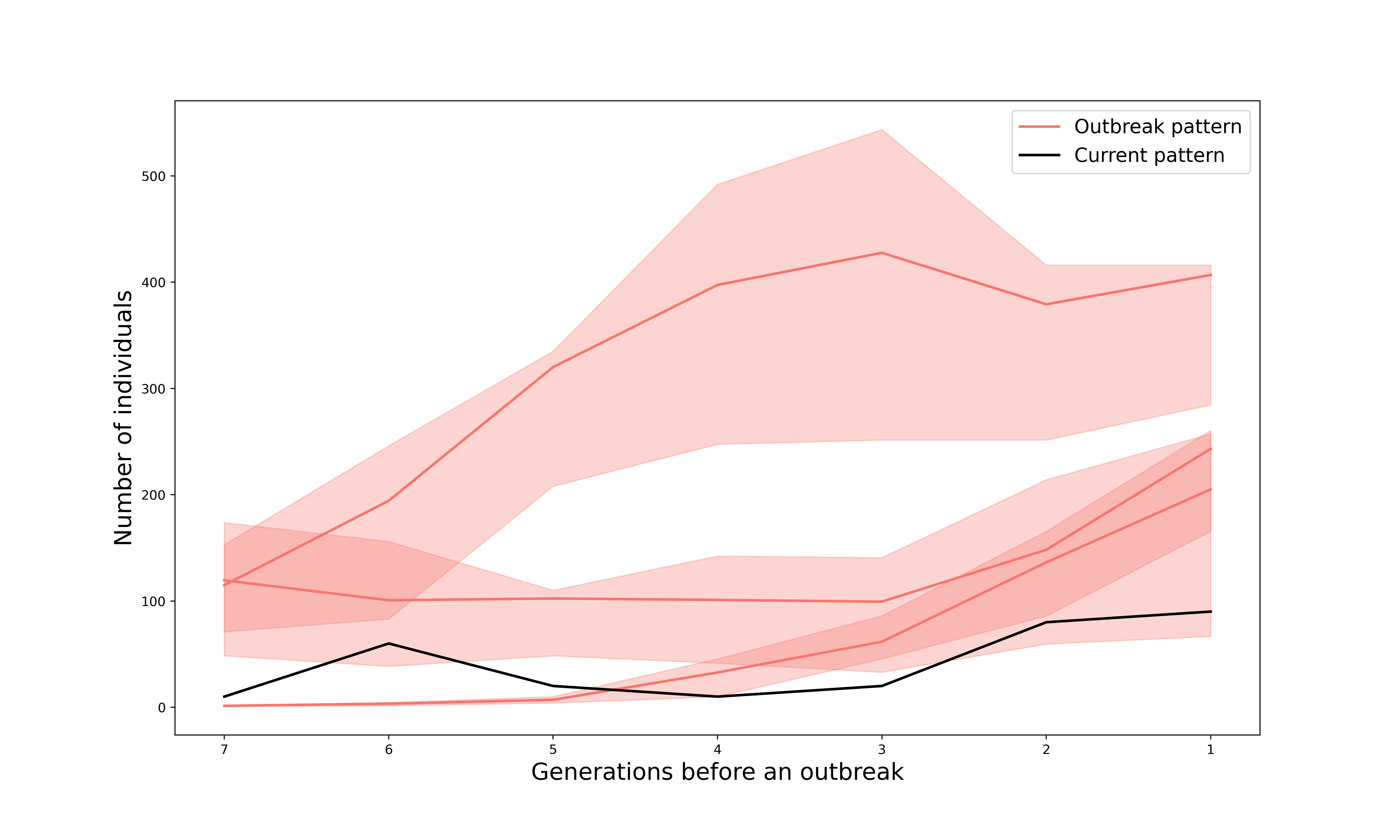}		
	\label{pypbpplot2}	
\end{figure}
By default, the optimisation is performed using 5-fold cross-validation by carrying out the GSA method with the negative area under the ROC curve as the objective function to be minimised, without pre-processing the time series using empirical mode decomposition. The function \texttt{pbp\_fit} contains other arguments, such as \texttt{verbose}, which prints the area under the ROC curve at every iteration of the optimisation process, and \texttt{maxfun}, which sets the number of evaluations of the objective function by the optimisation algorithm (GSA as default).

The object \texttt{clustered\_patterns} contains the cluster matrices estimated using the time series presented, the object \texttt{results} contains a data frame with the metrics obtained from the model: accuracy, fI-Score, precision, recall, true positive rate (TPR), false positive rate (FPR) and the estimated area under the ROC curve. It also contains the estimates for $m$, $d^*_{\mbox{\scriptsize cluster}}$ and $\alpha$. These metrics are presented for each of the four criteria used in our methodology to select $d^*_{\mbox{\scriptsize base}}$ (maximum FPR of 0.1 and 0.2 and minimum TPR of 0.8 and 0.9). The \texttt{parameters} object contains the estimates for $m$, $d^*_{\mbox{\scriptsize cluster}}$ and $\alpha$ in a dictionary ready for use. The function \texttt{pbp\_plot} can be used to visualise the obtained patterns. Finally, the details of each function used in the \texttt{pypbp} package are presented on the website \url{https://pypbp-documentation.readthedocs.io/en/latest/}.

\section{Pattern clustering algorithm \label{algorithm}}

Algorithm~\ref{algorithm} starts with pattern $\textbf{p}_{1}$, which represents the first row of the matrix $\textbf{P}$. We remove $\textbf{p}_{1}$ from $\textbf{P}$ and add it as the first row of $\textbf{P}^{\prime}_{1}$. After that we compute the association metric between $\textbf{p}_1$ and all subsequent rows of $\textbf{P}$. If $d(\textbf{p}_1, \textbf{p}_{j}) \geq d^*_{\mbox{\scriptsize cluster}}$, we add pattern $\textbf{p}_{j}$ as the last row of the cluster matrix $\textbf{P}^{\prime}_{1}$ and delete it from $\textbf{P}$. We repeat this process to obtain the cluster matrices $\textbf{P}^{\prime}_{c}, c=1,\ldots,C\leq I,$ until there are no more rows left in $\textbf{P}$.

\begin{algorithm}
	\SetAlgoLined
	\textbf{Input}: $d^{*}_{\mbox{\scriptsize cluster}}$ and $\textbf{P}$\\
	set $c = 1$\\
	set $l =$ number of rows of $\mathbf{P}$\\
	\For{$i$ \textbf{in} $\{1,2,\ldots,l\}$}{
		add $\textbf{p}_i$ as the first row of $\mathbf{P}_c^\prime$ and delete it from $\mathbf{P}$\\
		set $l^\prime =$ number of rows of $\textbf{P}^{\prime}_{c}$\\
		\For{$j$ \textbf{in} $\{1,2,\ldots,(I-\sum\limits_{c}l^\prime_c)\}$}{
			\If{$d(\mathbf{p}_i,\mathbf{p}_j)\geq d^*_{\mbox{\scriptsize cluster}}$}{
				append $\textbf{p}_{j}$ as the last row of $\textbf{P}^{\prime}_{c}$ and delete $\textbf{p}_{j}$ from $\textbf{P}$}
			update $l^\prime =$ number of rows of $\textbf{P}^{\prime}_{c}$}
		update $c = c +1$\\
		update $l = $ number of remaining rows of $\mathbf{P}$}
	\textbf{Output}: cluster matrices $\textbf{P}^{\prime}_{c}, c=1,\ldots,C\leq I$
	\caption{Obtaining cluster matrices $\textbf{P}^{\prime}_{c}, c=1,\ldots,C$, from $\mathbf{P}$.}
	\label{alg:cluster}
\end{algorithm}

\section{Comparison between optimisation methods \label{optimisation_methods}}

\begin{figure}[htb]
	\centering
	
	\includegraphics[width=17cm]{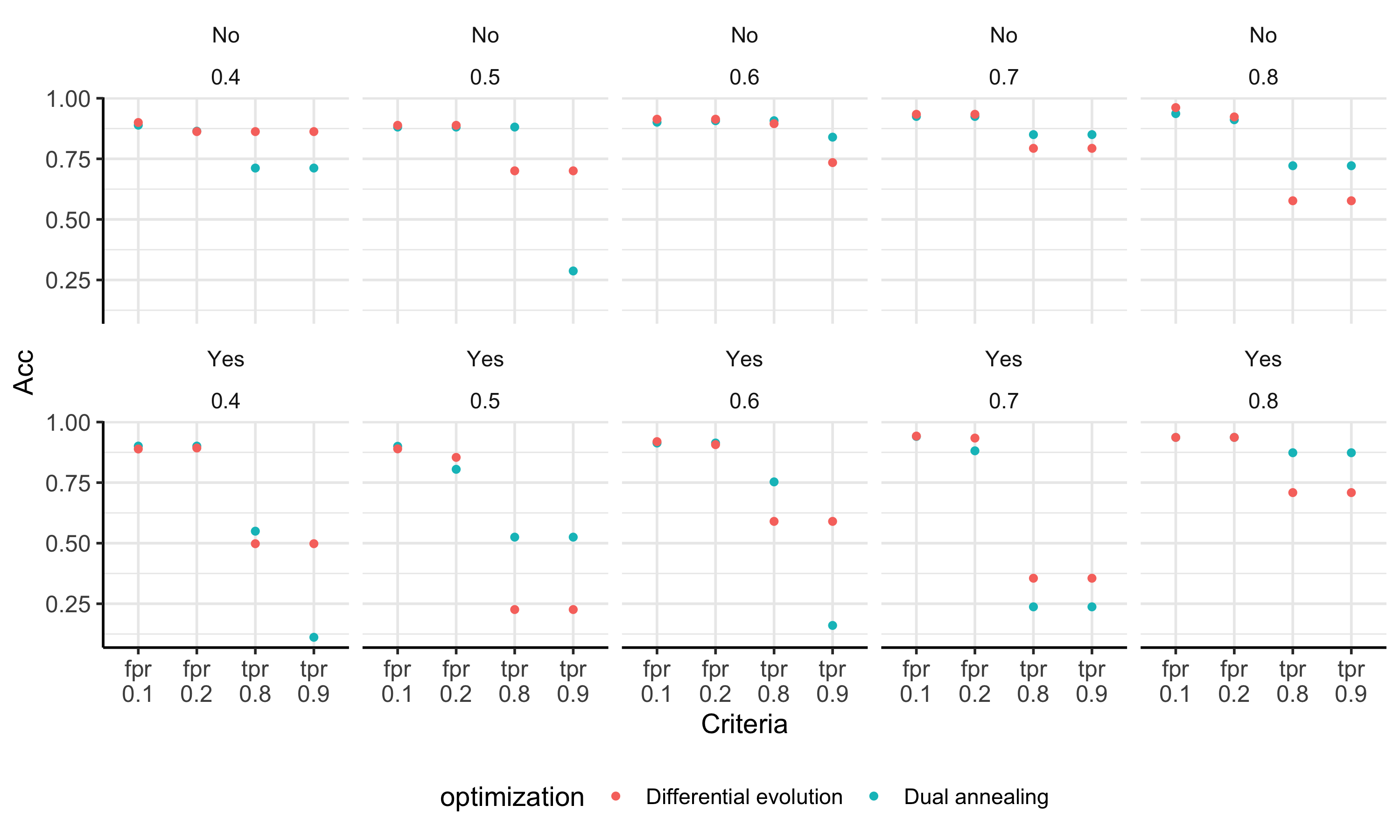}
	\caption{Accuracy obtained on the testing step of the PBP method. In both algorithms the following methods to choose $d^*_{\mbox{\scriptsize base}}$ were used: based on minimum FPR (.1 and .2 as minimum) and TFR maximum (.8 and .9 as maximum).}
	\label{optimization_acc}
	
\end{figure}

\begin{figure}[htb]
	\centering
	
	\includegraphics[width=17cm]{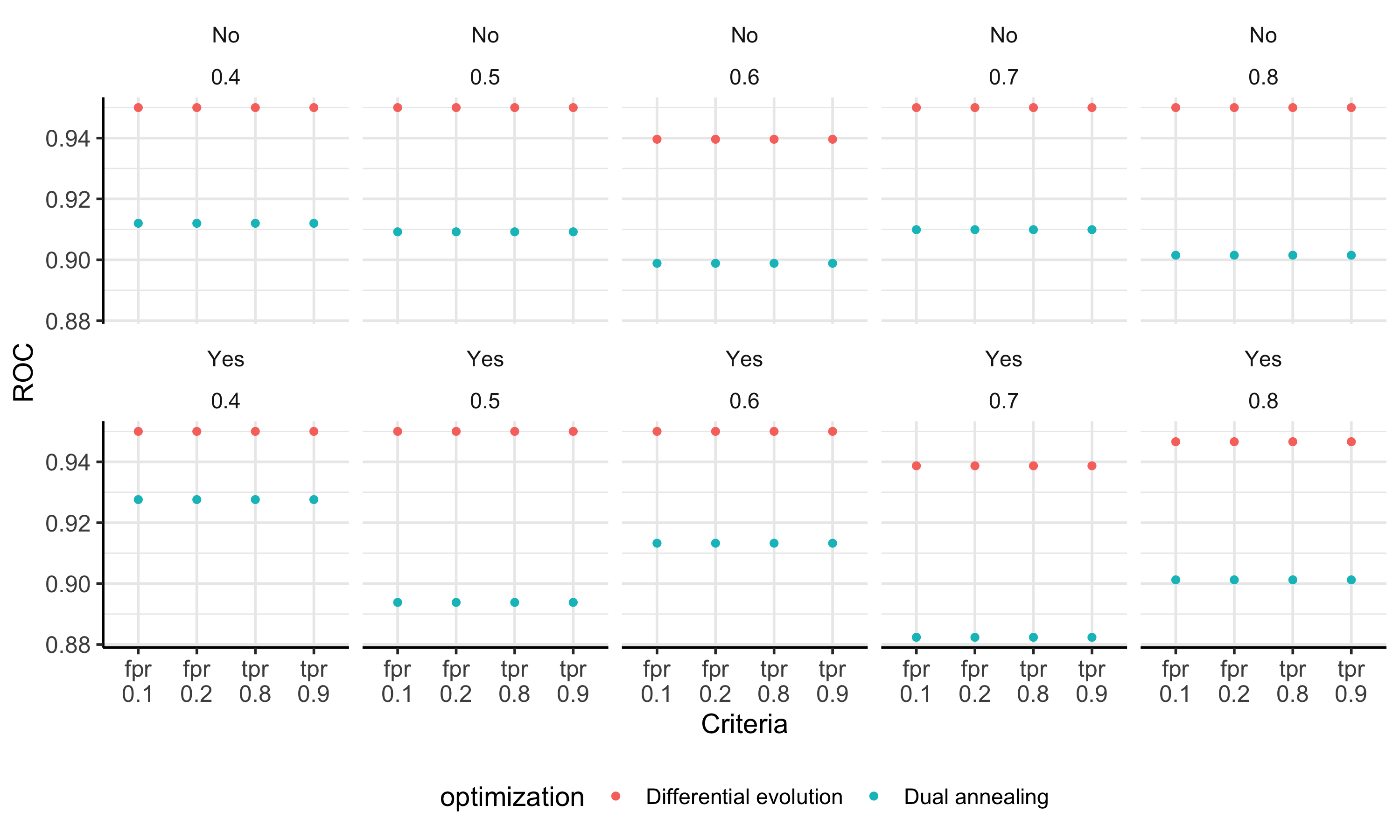}
	\caption{Area below the ROC curve obtained on the training step of the PBP method. In both algorithms the following methods to choose $d^*_{\mbox{\scriptsize base}}$ were used: based on minimum FPR (0.1 and 0.2 as minimum) and TFR maximum (0.8 and 0.9 as maximum).}
	\label{optimization_roc}
	
\end{figure}

\begin{figure}[htb]
	\centering
	
	\includegraphics[width=17cm]{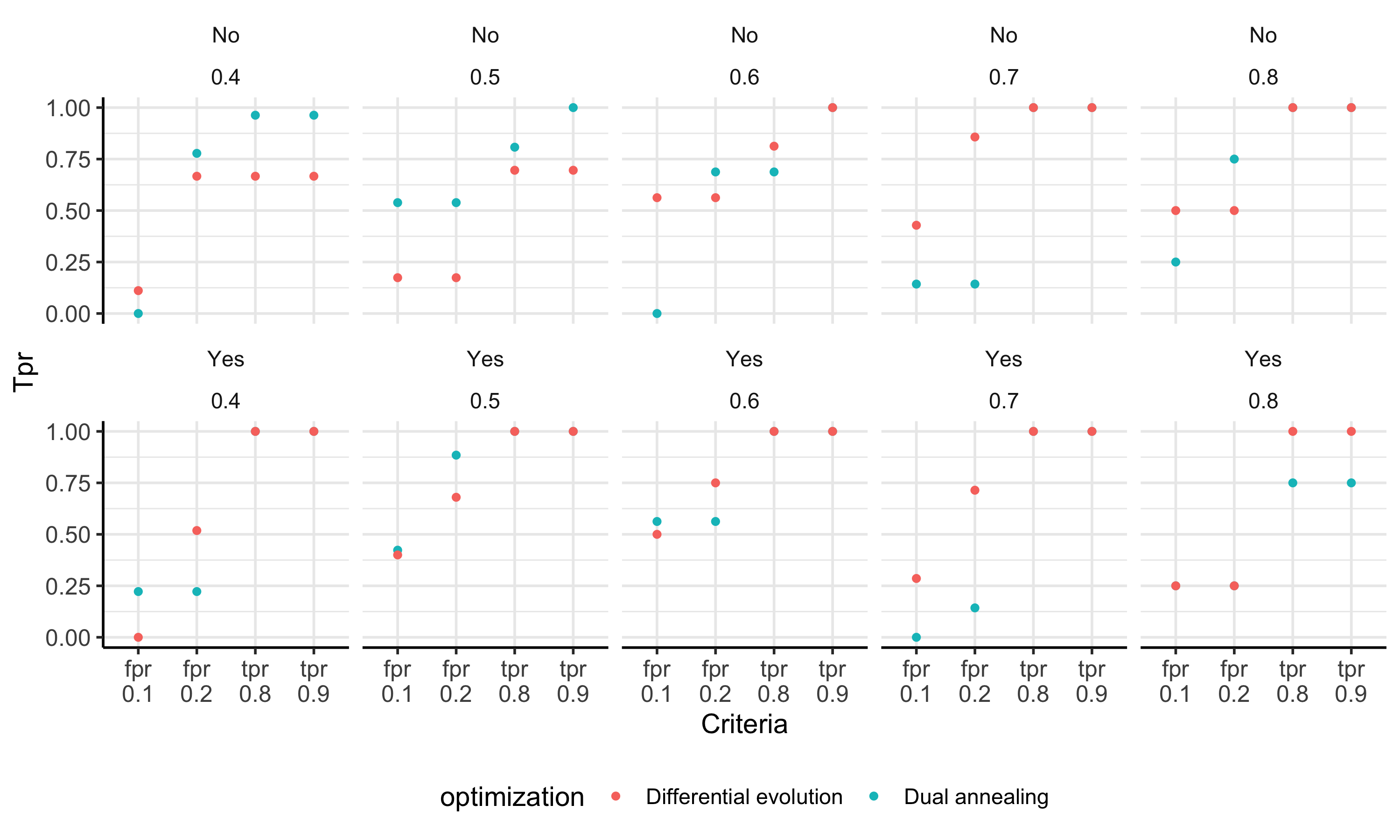}
	\caption{True Positive Rate obtained on the testing step of the PBP method. In both algorithms the following methods to choose $d^*_{\mbox{\scriptsize base}}$ were used: based on minimum FPR (0.1 and 0.2 as minimum) and TFR maximum (0.8 and 0.9 as maximum).}
	\label{optimization_tpr}
	
\end{figure}

\begin{figure}[htb]
	\centering
	
	\includegraphics[width=17cm]{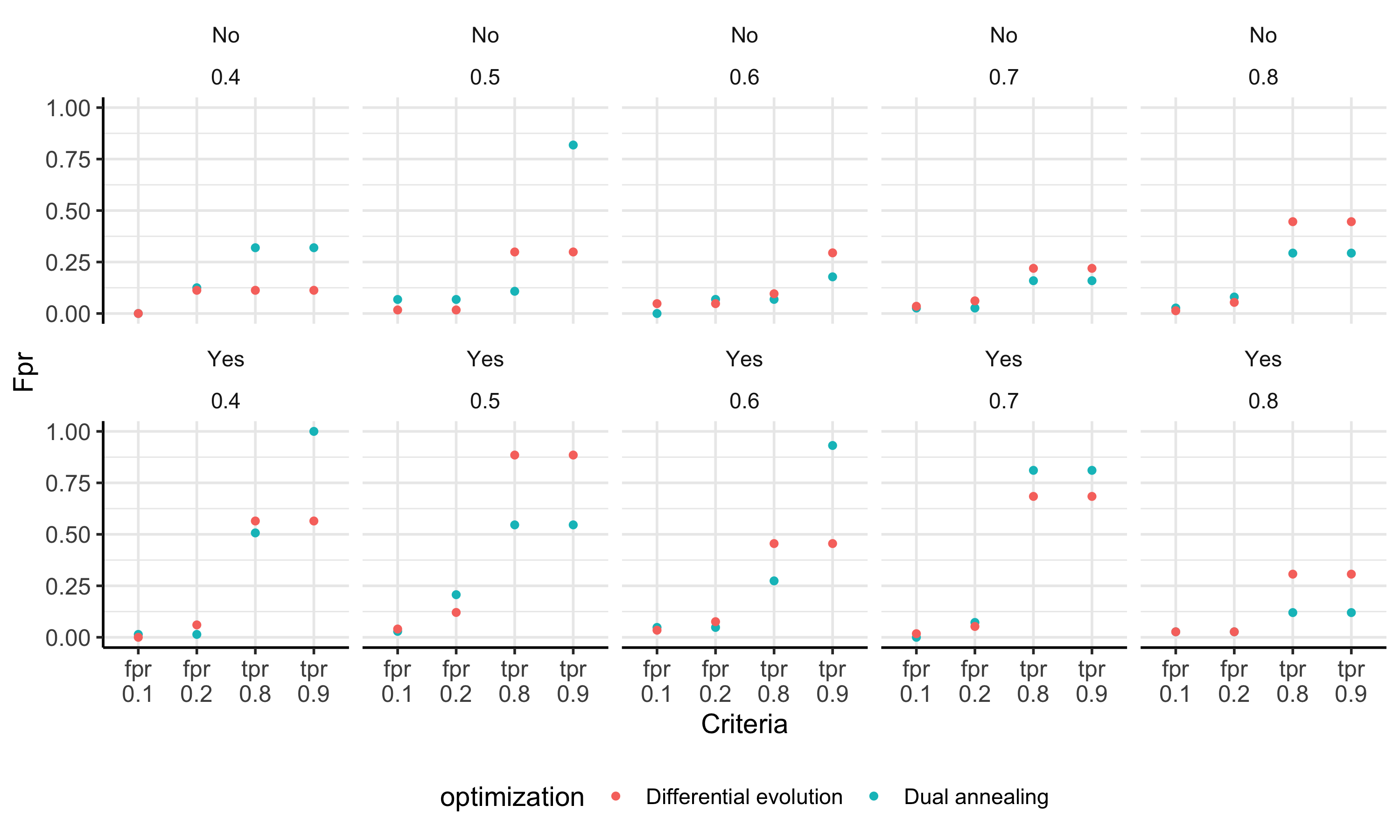}
	\caption{False Positive Rate obtained on the testing step of the PBP method. In both algorithms the following methods to choose $d^*_{\mbox{\scriptsize base}}$ were used: based on minimum FPR (0.1 and 0.2 as minimum) and TFR maximum (0.8 and 0.9 as maximum).}
	\label{optimization_fpr}
	
\end{figure}

\begin{figure}[htb]
	\centering
	
	\includegraphics[width=17cm]{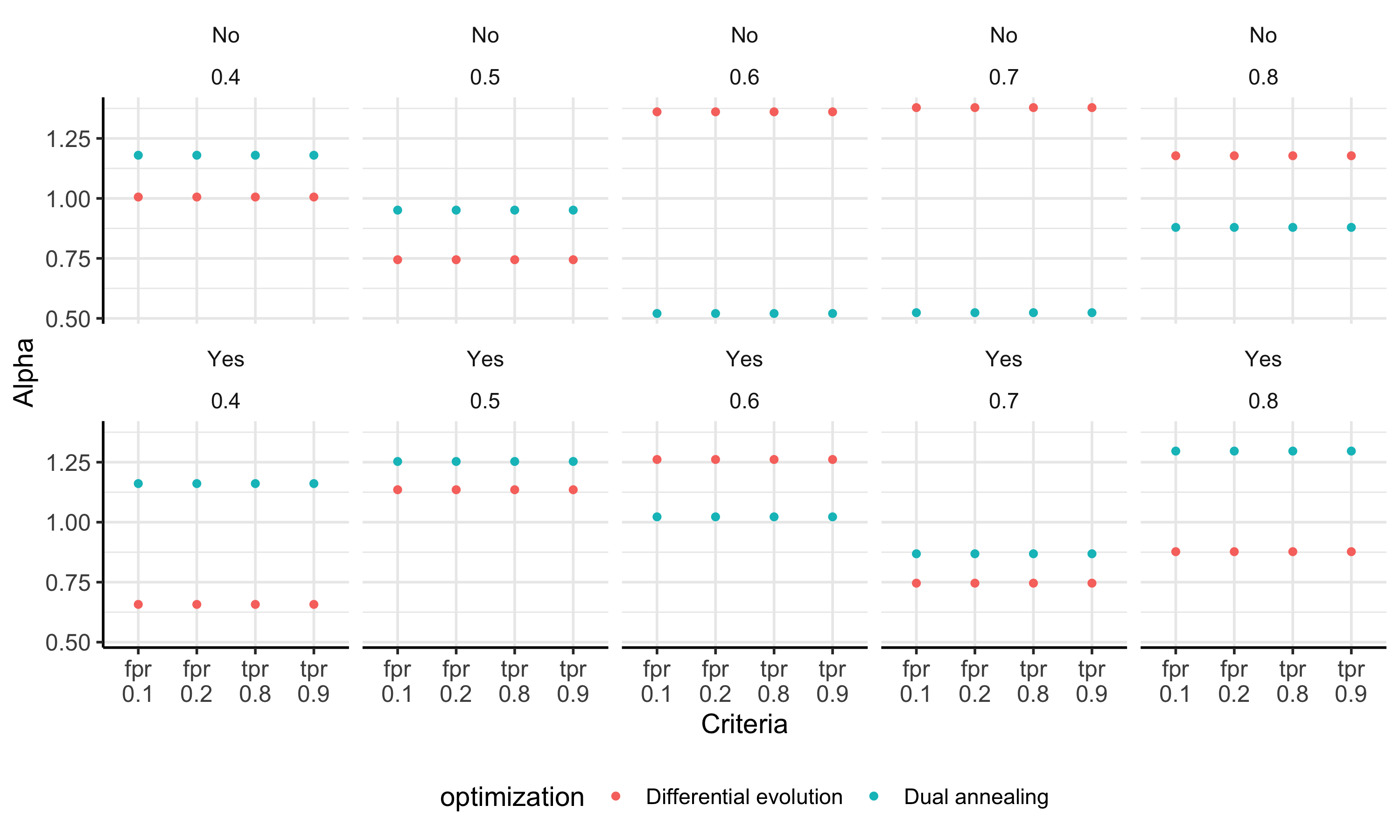}
	\caption{Estimated parameter $\alpha$ obtained on the training step of the PBP method. In both algorithms the following methods to choose $d^*_{\mbox{\scriptsize base}}$ were used: based on minimum FPR (0.1 and 0.2 as minimum) and TFR maximum (0.8 and 0.9 as maximum).}
	\label{optimization_alpha}
	
\end{figure}

\begin{figure}[htb]
	\centering
	
	\includegraphics[width=17cm]{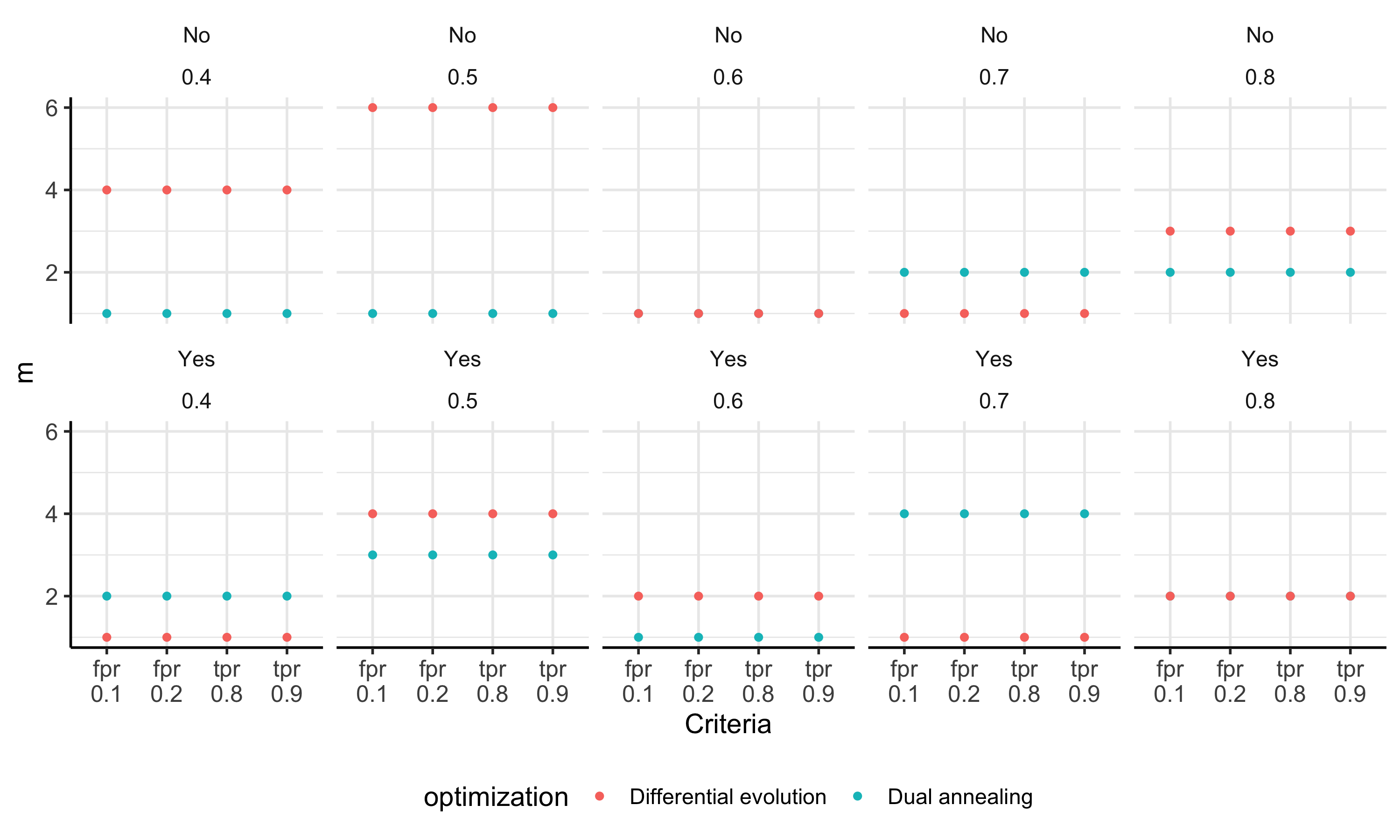}
	\caption{Estimated parameter $m$ obtained on the training step of the PBP method. In both algorithms the following methods to choose $d^*_{\mbox{\scriptsize base}}$ were used: based on minimum FPR (0.1 and 0.2 as minimum) and TFR maximum (0.8 and 0.9 as maximum).}
	\label{optimization_m}
	
\end{figure}

\begin{figure}[htb]
	\centering
	
	\includegraphics[width=17cm]{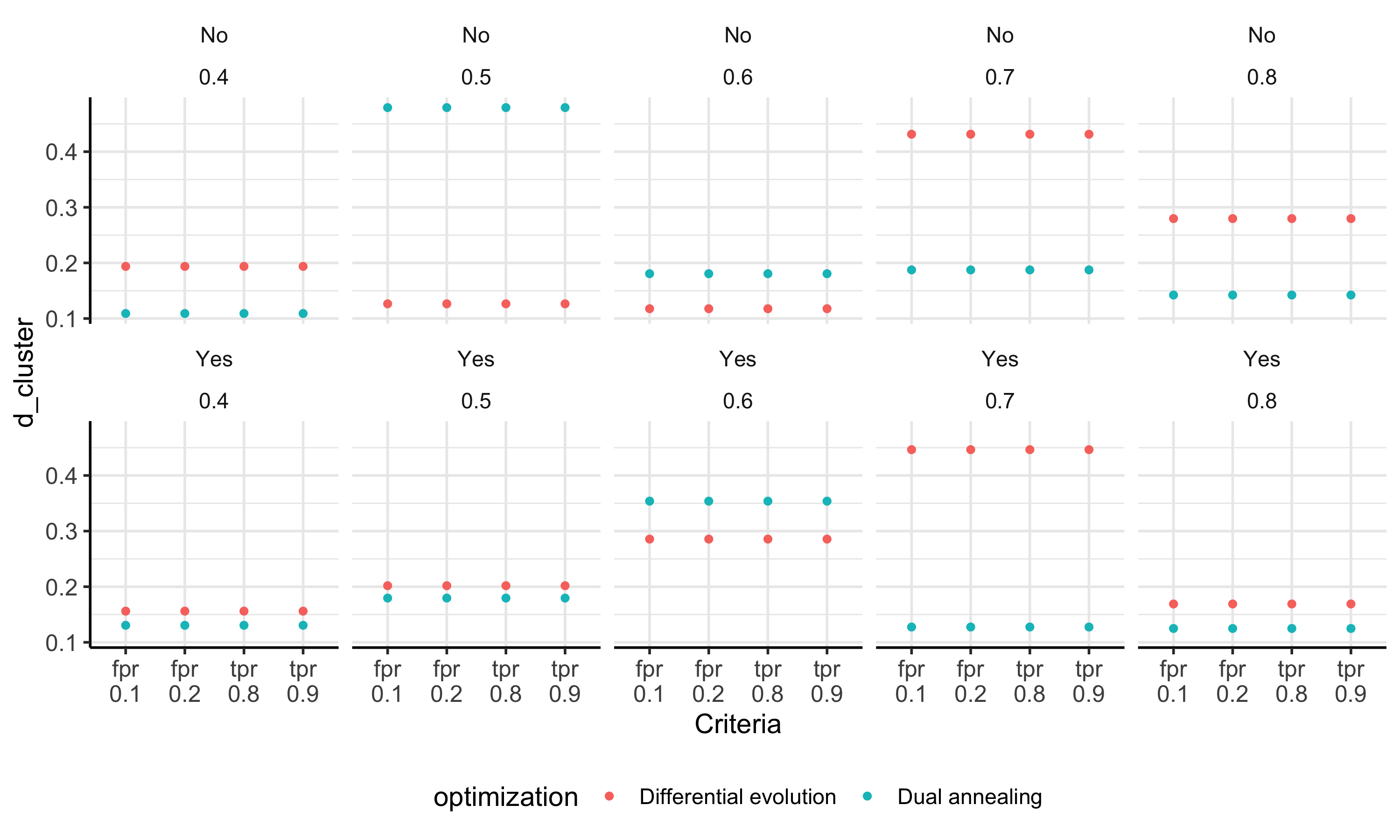}
	\caption{Estimated parameter $d^*_{\mbox{\scriptsize cluster}}$ obtained on the training step of the PBP method. In both algorithms the following methods to choose $d^*_{\mbox{\scriptsize base}}$ were used: based on minimum FPR (0.1 and 0.2 as minimum) and TFR maximum (0.8 and 0.9 as maximum).}
	\label{optimization_d_cluster}
	
\end{figure}

Here, we compare the differential evolution \citep{storn1997} and the  method \citep{tsallis1996} algorithms used to estimate $\alpha$, $d^*_{\mbox{\scriptsize cluster}}$ and $m$. We use these methods to optimise the area under the ROC curve (AUROC). We looked at the performance of our method using the time series data on aphids and parasitoids collected in Coxilha (Brazil-São Paulo). In Figure~\ref{optimization_acc} we present the accuracy, in Figure~\ref{optimization_roc} the area below the ROC curve (AUROC), in Figures~\ref{optimization_tpr} and~\ref{optimization_fpr} we present the true and false positive rates. In Figures~\ref{optimization_alpha}, \ref{optimization_m}, and~\ref{optimization_d_cluster} we present the estimated $\alpha$, $m$ and $d^*_{\mbox{\scriptsize cluster}}$ hyperparameters. 

Differential evolution is an algorithm based on evolutionary computation, which briefly consists of setting a population of candidate solutions and creating new updated candidate solutions based on the existing based on the score. For more detail see \cite{storn1997}. On the other hand, the generalised simulated annealing is a stochastic algorithm used for finding the global minimum of a given function in a continuous $D$-dimensional space. The method utilises a generalised entropic form
$$
    S_q = k\frac{1- \sum_i p^{q}_i}{q-1},
$$
where $q\in \mathbb{R}$, $p_i$ are probabilities of the microscopic configurations and $k$ is a conventional positive constant \citep{tsallis1996}. Based on this entropy \cite{tsallis1996} generalised its formulation to include the cases of the Boltzmann Cauchy machines, allowing to find local minima according to an acceptance temperature set in the generalised metropolis algorithm. 
Differential evolution was approximately ten times slower than generalised  method. This fact drove our decision to choose the  algorithm as the default algorithm for optimising the hyperparameters involved in the PBP method. However, users are allowed to choose which method to use when utilising the \texttt{pypbp} package.

\end{document}